\documentclass[12pt,a4paper,final]{iopart}

\usepackage{graphicx}
\usepackage[breaklinks=true,colorlinks=true,linkcolor=blue,urlcolor=blue,citecolor=blue]{hyperref}

\expandafter\let\csname equation*\endcsname\relax
\expandafter\let\csname endequation*\endcsname\relax
\usepackage{amsmath}

\begin{document}

\title[Wealth dynamics in a sentiment-driven market]{Wealth dynamics in a sentiment-driven market}

\author{Mikhail Goykhman}
\address{Enrico Fermi Institute, University of Chicago,\\
5620 S. Ellis Av., Chicago, IL 60637, USA}
\eads{\mailto{goykhman@uchicago.edu}}

\begin{abstract}

We study dynamics of a simulated world with stock and money,
driven by the externally given processes which we refer to as sentiments.
The considered sentiments
influence the buy/sell stock trading attitude,
the perceived price uncertainty, and the trading intensity of all or a part
of the market participants. 
We study how the wealth of market participants 
evolves in time in such an environment.
We discuss the opposite perspective in which the parameters of the sentiment
processes can be inferred a posteriori from the observed market behavior.

\end{abstract}


\section{Introduction}

Simulation is a possible way to approach the problem of modeling the
properties of a system with many degrees of freedom
and a complicated interaction pattern. In the cases when an analytical
description of a system is impossible one can try to
come up with a simulation based on a few built-in assumptions
in an attempt to model some of the prominent experimentally
observed phenomena.
In this spirit a simulated stock market models have been extensively studied in the literature,
with the goal to describe the real stock price behavior and investigate strategies of market participants.
Some of the original models have been proposed
in \cite{PalmerAH1994,ArthurHL1996,PalmerAH1998,LeBarronAP1999,Raberto2000,Bonabeau2002,PontaRC2011,
Bertella2014},
for a review see \cite{SamanidouZS2007}, and references therein.

A typical simulated market
environment includes a large number of agents possessing units of cash
and shares of stock (the simplest models consider the world with just one
kind of stock, although multi-asset models exist, in particular among the references above), who manifest their trading activity by submitting stock buy or sell orders
to the stock exchange. One possibility is to have a stock exchange which
puts orders into the order book,
and fills orders
by searching for an intersection of the stock supply and demand curves,
determining the equilibrium stock price, and clearing the possible buy and sell orders.
This is re-iterated over a large number
of steps and the resulting emergent stock price time series is observed.

Generally the artificial stock market models consider agents which perform
their trades in accord with certain strategies, such as 
trend-following, contrarian, fundamental trader, etc. That is, the agents might be basing
their strategy on analyzing and drawing conclusion from the past stock price behavior. Various
groups of agents might be set up to compete against each other,
and the winning trading strategy is optimized in real time by the agents,
as the new information becomes available to them (see {\it e.g.} \cite{LeBarronAP1999} for the
model of agents using genetic algorithms to shape the most optimal trading strategy).
Such models are known to have successfully reproduced some of the
well-recognized facts about the stock prices time series, such as
fat tails of the logarithmic stock returns \cite{Mandelbrot1963} and volatility
clustering.

\vspace{0.2cm}

In contrast to the models reviewed above, the simulated market environment
which we set up and implement in this paper does not endow its agents with
any ability to analyze information from the past stock time series, or any mechanism
to make predictions of a possible future directions of the stock price.
The agents therefore will not attempt to readjust their portfolios for
any particular optimization purpose.
Instead, behavior of agents in our model will be influenced
by the market environment driven
exclusively by an externally given processes,
which we call sentiments. (A different kind of a sentiment incorporation
into the agents's behavior can be found in \cite{PastorePC2010}.) Agents do not influence each other's sentiment, but all (or large groups)
of agents receive the sentiment from the same source. 

In other words, we will be using the framework in which
the state of the market, considered in the given
time period, is uniquely defined by the sentiment
processes driving the market evolution during that period.
In this framework we assume that the strategies of market
participants converge collectively to what 
can be effectively modeled by a driving sentiment process.
The choice of a specific sentiment process
will constitute our prior belief about the stock market.
For instance, all of the agents might settle into a belief
that arrival of a breaking news about the stock means the volatility surge by {\it e.g.} factor of two.
Or a subgroup of agents might decide that the stock is overvalued, and be
more willing to sell it than buy it.
The task is then to determine how the stock price is going to behave in the market where
the participants follow these attitudes. This paper is concerned with such a task.\footnote{
We can attempt to go further and determine posterior probabilities for various possible sentiment
processes considered to be driving the market. It would be interesting to develop
a full Bayesian framework to fulfill this task. We discuss this in section \ref{discussion}.}

\vspace{0.2cm}

At each time step $t$ the first question which faces the agent is whether to
stay out of the market during the step or submit some order.
We will model the market activity by our agents as the Poisson
process, with the agent's orders being
 separated by exponentially distributed waiting times.
  (In \cite{EngleR1997,EngleR1998,PontaSR2011} it was pointed
out that in reality inter-trade separation times follow a Weibull distribution.)
The mean waiting time between the trades, $\rho(t)$,
is itself a random process. It is defined as an external trading intensity
sentiment, which is used by all or a part of the agents to gauge intensity of their own
market activity. Each agent $A_n$ picks its own $\rho _n(t)$ as a gaussian
draw around the commonly given $\rho(t)$.

If the agent decides to participate in the market during the given
trading session it needs to decide whether it wants to buy or sell the stock.
We specify that this decision is also influenced by an external
sentiment $\psi(t)$, common for all (or large groups of) agents. The $\psi_n(t)$ for the
agent $A_n$ is a gaussian random variable, centered around $\psi(t)$,
and prescribing the buy vs sell dis-balance sentiment. When $\psi_n=0$
the agent is equally likely to buy or sell, it leans towards buying when $\psi_n>0$,
and towards selling when $\psi_n<0$, as we describe in section \ref{The_disbalance_sentiment}.

Once the agent decides to buy or sell it needs to determine the limit
price for its order. We prescribe that regardless of the buy or
sell side of the order the agent will draw the limit price as a gaussian random
variable with the mean equal to the most recent stock price, and the standard
deviation $\sigma_n(t)$ defined by the external volatility sentiment $\sigma(t)$.

The $\sigma(t)$ is an external time-series process, which we specify as the Poisson
process of the jump volatility kind \cite{Cox1976,Merton1976}. For simplicity we assume that the volatility
$\sigma(t)$ takes the calm value ${\cal N}(\sigma_c,\delta\sigma_c)$
most of the time, and the breaking news value ${\cal N}(\sigma_b,\delta\sigma_b)$,
arriving once in a while according to the exponential distribution with the mean $\lambda$.
The $\sigma_b$ is significantly larger than $\sigma_c$, and the non-vanishing small $\delta\sigma_{c,b}$
are introduced for gaussian randomization purposes.

The last needed ingredient is given by the size of the order.
We can augment our model by giving the agents some ability to 
plan the strategies, deciding on what specific size of the order to submit.
However we leave this for the future work, and refrain to a uniform
draw \cite{Raberto2000} for the order size for the rest of this paper.

\vspace{0.2cm}

We use our market environment described above to explore dynamics
of wealth in the simulated society of agents.
We will consider various starting allocation distributions for the wealth
of the agents: identical, uniform, gaussian, and Pareto. We discover
that regardless of the initial wealth distribution 
the resulting distribution quickly converges (in the tail of 25\% of the
wealthiest participants) to the Pareto
law, consistent with the analogous studies in the literature
\cite{Bouchaud2000,HuangS2001,RabertoCF2003}.

For recent empirical studies of the wealth power-law distribution
see \cite{LevyS1997}. See also \cite{Milakovic2003,CastaldiM2007}, which applied
the statistical equilibrium ideas of \cite{Foley1994} and the maximum entropy distribution principle of
\cite{Jaynes1957} to argue in favor of the power law distribution of wealth,
see \cite{Lux2008} for a review.

The rest of this paper is organized as follows.
In section \ref{Market_environment} we set up the market
environment in the most general form, preparing
the background for the simulation models which we will
be studying in the subsequent sections. In section \ref{Pareto_wealth_distribution}
we review some of the relevant properties of the Pareto distribution,
which will be useful for our analysis of the wealth distribution
in the society of agents.
In section \ref{section_4} we study dynamics of the wealth distribution
in a simple simulation over a large number of steps. 
In section \ref{section_5} we study the jump volatility model
with a non-trivial buy/sell sentiment process, and investigate
the resulting stock price behavior. In section \ref{section_6}
we separate the system into four subgroups of agents, each receiving
its own sentiment, and study the resulting wealth dynamics.
We discuss our results in section \ref{discussion}. Section \ref{The_matching_engine}
is an appendix where
we describe design of our stock exchange.

\section{Setting up the market environment}\label{Market_environment}

In this paper we study dynamics in the simulated market environment
which is provided by the market participants (agents) and
the stock exchange.
The stock exchange mediates interaction between the agents
by facilitating trading of shares of one kind of stock.
Additional details might be imposed
as to how the agents are implementing their market activity, and the specific parameters
will be considered in the subsequent sections. In this section we describe the most
general setting which will be the core for all of the models considered in this paper.
We begin by describing the trading environment and the market participants,
and then outline the sentiment processes which will be driving
dynamics in our models.

We will be studying
dynamics of the market environment in a discrete time $t=1,\dots,T$,
where $T$ is the total number of simulation steps.
We will be considering the system of $N$ agents, each
attributed with a portfolio of $m$ units of cash (non-negative real-valued) and $s$ shares of stock (
non-negative integer-valued).
The basic act of market activity of the agent is defined as
submitting the buy or sell order to the stock
exchange. The agent possessing $m$ units of cash might decide to use any part
of that amount to submit an order to purchase shares of stock. The agent has to decide
on the price for which it is ready to make the purchase. If it chooses the price $p$,
then it can order to buy anywhere between one and $[m/p]$ shares of stock, where square
brackets stand for an integer part. Similarly, if the agent possesses $s$ shares of stock it can decide
to sell any part of that number for a certain price.

The stock exchange receives and fills orders by maintaining the order book
and operating the matching engine. The order book contains
 tables of buy and sell orders, sorted by their price, and specifying
the sizes of the orders.
At each time step the orders of all of the agents are first recorded into the order book.
When the stock exchange attempts to fill the orders
currently present in the order book it uses the matching engine,
which constructs the cumulative sell and buy orders (supply and demand curves)
and searches for their intersection. If $P_*$ is the intersection price,
the matching engine then fills the orders for the clients: the buy orders
at the price $P_*$ and higher, and the sell orders at the price $P_*$ and lower.
The filled orders are deleted from the order book by the matching engine.
The matching engine also distributes the trade proceeds to the clients.
Such a set-up is typical for the stock market simulations, see for instance \cite{Raberto2000}.
Depending on the specifics of the considered model we might
decide to keep the un-filled orders in the order book, or clear up the order book
before the next step. In the simulations considered in this paper at the end of each step
all the remaining orders from the order book are removed.
Details of the stock exchange structure are given in section \ref{The_matching_engine}.

At the beginning of each time step, before
getting the possibility to submit an order to the stock exchange,
each agent receives the interest rate return $r$ on its cash and the dividend yield $d$
on its stock. If by the end of step $i-1$ the agent has $m$ units
of cash and $s$ shares of stock, and the stock price is $P_{i-1}$, then at the beginning of step $i$
the agent will have $(1+r)m+(1+d)P_{i-1}s$ units of cash and $s$ shares of stock, and it can
use that money and stock to participate in the $i$th trading session. To be specific,
we will consider the fixed interest rate and the gaussian-distributed
(around some positive mean) dividend yield $d$, these settings
are similar to the known models in the literature, {\it e.g.} \cite{PalmerAH1994,LeBarronAP1999}.
The precise numbers will be chosen for the specific models considered below.

Existence of a non-zero interest rate and dividend yield implies that
our system is not closed: the agents keep their cash in the bank (which is located
outside of the system) which pays them the
interest. The agents also interact with the firm (which is also located outside of the system) whose stock they hold and who pays
up parts of the profits (which fluctuate in time) to the shareholders.
The increased money supply available to the agents will in turn push up the stock
price, unless we explicitly put in the condition that the agents are curbing their demand for stock,
or optimizing it by some considerations, for instance  \cite{Markowitz1952}.

Our model can be generalized by incorporating purposeful trading strategies
for the agents who will try to maximize their wealth, for instance through seeking the dividend
return on stocks rather than the interest return on cash, when they
predict the former to be higher than the latter, or vice versa. We leave this
sophistication for future work.
In the context of this paper non-vanishing $r$ and $d$ will mostly
result in a simple inflation of the stock price, which can be
discounted by an appropriate factor.
In the world with many assets, of course, the flux of money
to and from a particular investment vehicle is defined by a prescribed
strategy, evaluating its investment attractiveness,
and is far from being reduced to a simple price inflation.

\subsection{The buy/sell dis-balance sentiment}\label{The_disbalance_sentiment}

When $r=0$ and $d=0$, under simple enough
agent strategy assumptions, the stock
price will quickly reach an equilibrium value, around which it will be exhibiting
the mean-reverting fluctuating behavior.
Indeed, consider a simple
model where the traders at each time step submit a buy order with the probability $p_b$
on the fraction $x$ of its cash,
or a sell order with the probability $p_s$ on the fraction $x$ of its shares,
where in either case the $x$ is a random number uniformly distributed in $[0,1]$,
Suppose $S$ is the total number of shares outstanding, and $M$ is the total
cash of the agents. If $p_b=p_s$, the stock price will quickly reach the equilibrium value,
$P_e=M/S$, and then fluctuate around it.
If  $p_b\neq p_s$, the stock price would
again behave in the mean-reverting way, but around the new equilibrium point,
determined by the balance of average supply and demand flows,
\begin{equation}
P_e =\frac{p_bM}{p_sS}\,.\label{equilibrium_buy_sell}
\end{equation}
This can be easily confirmed by simulations.

While putting in a fixed dis-balance between the supply and demand probabilities $p_s$
and $p_b$ does not lead to a long-term trend, as pointed out above, it does result
in a transitional period between the old and the new equilibrium stock price points.
Changing the buy/sell dis-balance will activate a new short-term trend in the stock price.
Therefore we could construct a model in which a subgroup or all of the agents would
trade in a dis-balanced way, preferring buying over selling, or vice versa,
creating the short-term price trends.
We can formalize this by considering a driving sentiment process $\psi(t)$, such that
\begin{equation}
\frac{p_b}{p_s}=e^{\psi}\,,\label{buy_sell_disbalance}
\end{equation}
and $\psi$ is real-valued. (Earlier model incorporating the buy/sell
sentiment has been formulated in \cite{PastorePC2010}, although it essentially differs
from our model by the specific meaning assigned to the sentiment factor.)
Combining the expression (\ref{equilibrium_buy_sell})
for the equilibrium price with the cash dynamics $M(t)$ (from interest
and dividends), and the sentiment driver (\ref{buy_sell_disbalance}), we obtain
the time dependence of the equilibrium stock price
\begin{equation}
P_e(t) =\frac{M(t)}{S}\,e^{\psi(t)}\,.\label{equilibrium_psi_price}
\end{equation}

In the simplest model all or a large group of the agents will
gauge their $p_b/p_s$ sentiment by the same value of $\psi$ (specifically, drawing their
sentiment $\psi_n$ randomly from the gaussian with the mean $\psi$).
Such a model emulates a crowd-like behavior, driven by an external source.
Overall, the price of the stock can be pushed
this way pretty far up or down.

\subsection{The jump volatility}

Suppose at the step $i-1$ the price of the stock was $P_{i-1}$,
and at the step $i$ the agent wants to submit a buy or sell
order on the stock. The agent has to come up with the size of the order
(the number of shares to buy or sell) and the limit price for which it is ready
to have its order filled. We will be following the simple volatility sentiment assumption in which 
each agent $n$ decides on the limit price $P^{(n)}_i$ by drawing it from the gaussian
distribution with the mean $P_{i-1}$ and the variance $\sigma$ (regardless
of the buy or sell side of the submitted order, unlike \cite{Raberto2000}),
\begin{equation}
P^{(n)}_i=|{\cal N}(P_{i-1},\sigma)|\,,\label{limit_price_submitted}
\end{equation}
where we have taken an absolute value to ensure that
each agent submits a positive-valued price.
We notice right away that taking an absolute value 
results in a higher probabilistic weight for the lower price, thereby adding
a slight effective selling sentiment to the market. We confirm
this point by simulations, and notice that decreasing $\sigma$
alleviates this artifact.

We assume that all of the agents are following the same news source,
and use the information obtained from that news to
gauge their uncertainty about the stock price.
To be precise, we assume that the news can be in two states:
calm and breaking. When the news are calm the agents draw
$\sigma$ from the gaussian distribution with the mean $\sigma_c$,
and the variance $\delta\sigma_c$. When the news are breaking
the agents choose the $\sigma$ from the gaussian distribution with
the mean $\sigma_b$ and the variance $\delta\sigma_b$,
where $\sigma_b$ is significantly larger than $\sigma_c$. 
The breaking news arrive according to the Poisson process,
with the mean inter-arrival time $\lambda$.
The probability of no news at time $t$ since the last observation at time $t_0$ is
\begin{equation}
P(t|t_0)=e^{-\frac{t-t_0}{\lambda}}\,.\label{sigma_process}
\end{equation}
Such a volatility behavior is known in the literature as the jump volatility model
 \cite{Cox1976,Merton1976}.
We demonstrate that incorporating jump volatility into our simulated
market environment results in the fat tail distribution of the stock returns.

\subsection{The market participation}

One possible way to set up a stock market simulation would
be to make all of the agents submit random orders to the stock
exchange at each time step \cite{Raberto2000}. This assumption is not very
realistic, and unnecessary increases the time cost of the simulation by clogging 
up the order book.
It was proposed to model the agents submitting their orders
according to the Weibull/Poisson process, thereby ensuring a desirable average
fraction of agents participating in the market activity at each time step \cite{PontaSR2011}.
The probability of no trade by the agent $A_n$ at time $t$ since the last observation at time $\tau_n$ is
\begin{equation}
P_n(t|\tau_n)=e^{-\frac{t-\tau_n}{\rho}}\,.\label{rho_process}
\end{equation}
where $\rho$ is the mean inter-trading time. In general we want $\rho$ to be
a time-series process, feeding the trading intensity mood into the market.

\subsection{Summary of the market sentiment drivers}

To summarize, we propose to consider the market environment
in which the agents gauge their trading activity, the price position, and
the trading intensity attitude by reading the sentiments from the common
(for all or a subgroup of the agents)
external source. The sentiments are supplied in the form of three time series
processes:
\begin{itemize}
\item
The buy/sell dis-balance sentiment $\psi$ determines the 
bullish vs bearish attitude (\ref{buy_sell_disbalance})
and creates a short-term price trends.
\item
The volatility sentiment $\lambda$ influences the price
uncertainty $\sigma$ perceived by the market participants
through
the jump process (\ref{sigma_process}).
\item
The trading intensity sentiment $\rho$ drives the number
of market participants via the Poisson process (\ref{rho_process}).
\end{itemize}

\subsection{Initial wealth allocations}

For all of the simulations in the subsequent sections we will consider four possible
 initial wealth allocations:
identical, uniform, normal, and Pareto,
\begin{itemize}
\item
Identical, ${\cal W}_0={\cal N}(2\times 10^6,10^3)$, with a small gaussian variability.
\item
Uniform, ${\cal W}_0={\cal U}(0,10^6)$.
\item
Normal, ${\cal W}_0={\cal N}(10^6,3\times 10^5)$, with a large
gaussian variability.
\item
Pareto, ${\cal W}_0={\cal P}(10^5,1.5)$, where $w_0=10^5$, $a=1.5$, see (\ref{pareto_distribution}).
\end{itemize}

\section{Pareto wealth distribution}\label{Pareto_wealth_distribution}

It is known that the wealth distribution of agents
participating in a generic simulated market activity converges in the tail
to the Pareto distribution \cite{RabertoCF2003,PastorePC2010}. The results of this paper
agree with such a conclusion, regardless of the choice
of the initial wealth allocation.
Therefore it is useful to review some properties of the Pareto
distribution, which is the goal of this section.

The Pareto density function
is given by
\begin{equation}
f(w)=\frac{a w_0^a}{w^{1+a}}\,,\label{pareto_distribution}
\end{equation}
for the wealth $w\geq w_0$ and the Pareto exponent $a$
(the convention is that the actual exponent
in the p.d.f. (\ref{pareto_distribution}) is $1+a$).
If $n_0$ is the total number of agents with the Pareto-distributed wealth,
then the expected number of agents with the wealth not less than $w$ is
\begin{equation}
n(w)=n_0\,\int _w^{\infty} dw' \,f(w')=n_0\left(\frac{w_0}{w}\right)^a\,.
\label{rank_of_agent}
\end{equation}
We can view $n(w)$ as the rank of the agent, where the agents
are sorted according to their wealth, and ranked from rich to poor.
Therefore the Pareto exponent $a$ can be determined by fitting
the linear dependence of the logarithm of the rank on the logarithm
of the wealth \cite{LevyS1997}
\begin{equation}
\log w\sim -\frac{1}{a}\log n+c\,,\label{alpha_fit}
\end{equation}
for some constant $c$. 

Further on we can calculate the total
wealth of the agents, each of which has the wealth not lower than $w$,
\begin{equation}
W(w)=\int _w^\infty dn(w')w'
 =n_0\int _w^\infty dw'\,w' f(w')=\frac{a n_0w_0^\alpha}{(a-1)w^{a-1}}\,.
\end{equation}
Combining this with (\ref{rank_of_agent}) we can derive the cumulative
wealth of the richest $n$ agents
\begin{equation}
W(n)=W_0\left(\frac{n}{n_0}\right)^\alpha\,,\qquad W_0=\frac{n_0w_0}{\alpha}
\,,\qquad \alpha=\frac{a -1}{a}\,,
\end{equation}
where $W_0$ is the total wealth.
Then the wealth fraction $R=W/W_0$
held by the richest fraction $x=n/n_0$ of the agents is
\begin{equation}
R(x)=x^\alpha\,.\label{pareto_rule}
\end{equation}
The fraction law (\ref{pareto_rule}) results in the Pareto principle.
Specifically, the famous 80-20 principle, according to which
20\% of the richest agents hold 80\% of the total wealth,
will result from $\alpha=0.14$, that is, the Pareto exponent $a=1.16$.
Similarly, as another example, 1\% of the richest agents will hold 40\%
of the wealth when $\alpha=0.2$ and $a=1.25$.

\section{Society driven by a simple participation sentiment}\label{section_4}

\begin{figure}
 \includegraphics[width=8cm, height=5cm]{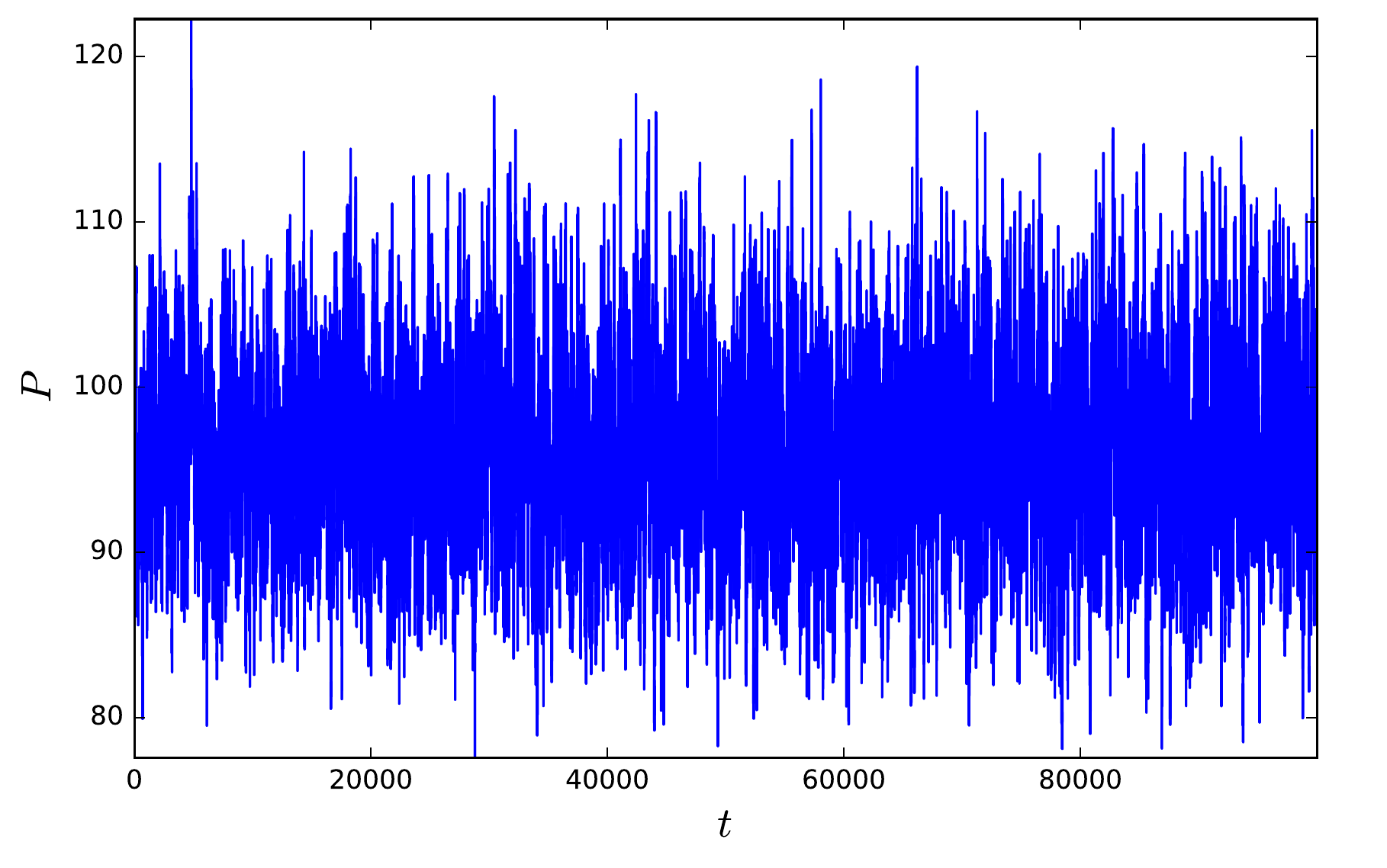}
  \includegraphics[width=8cm, height=5cm]{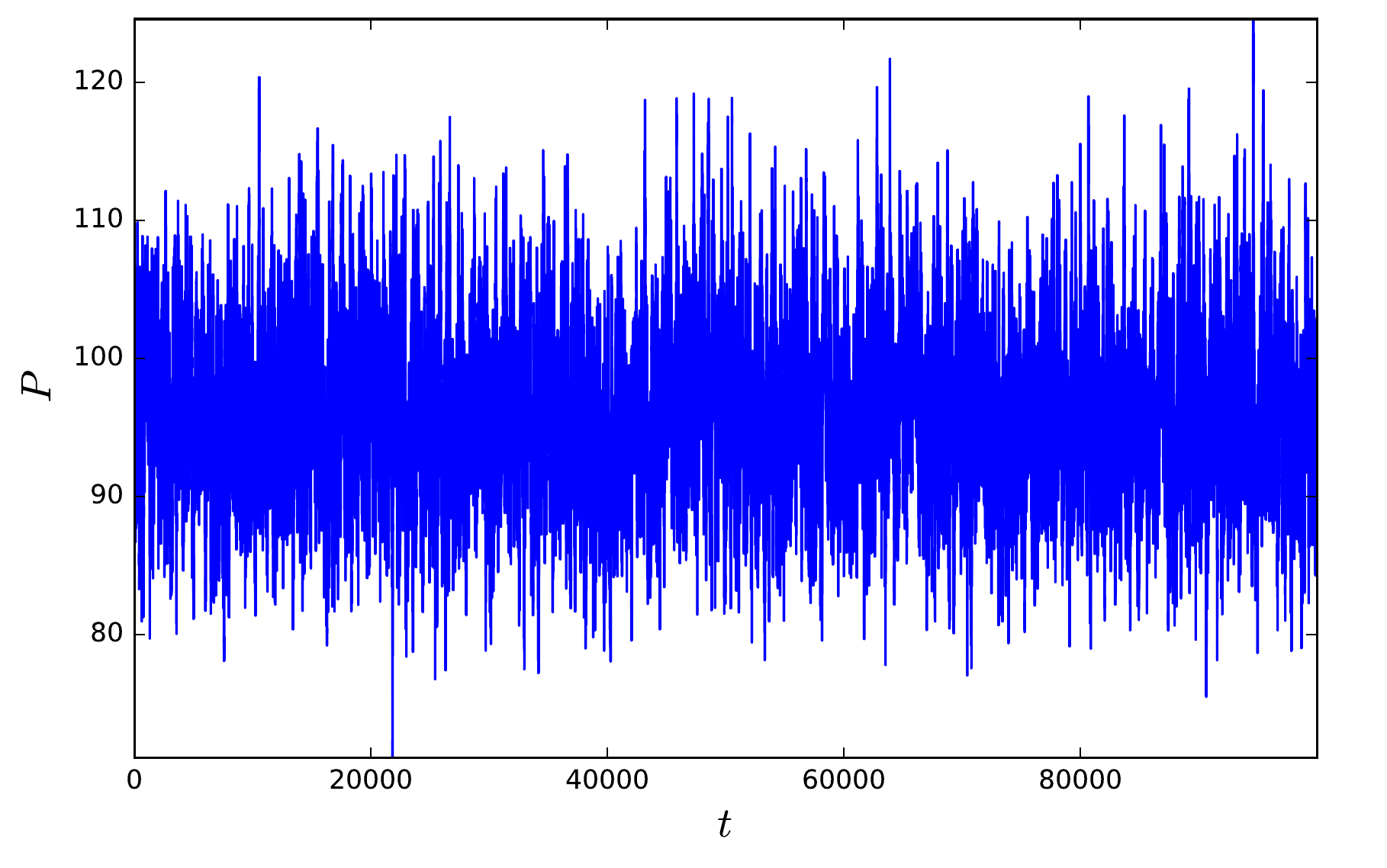}
   \includegraphics[width=8cm, height=5cm]{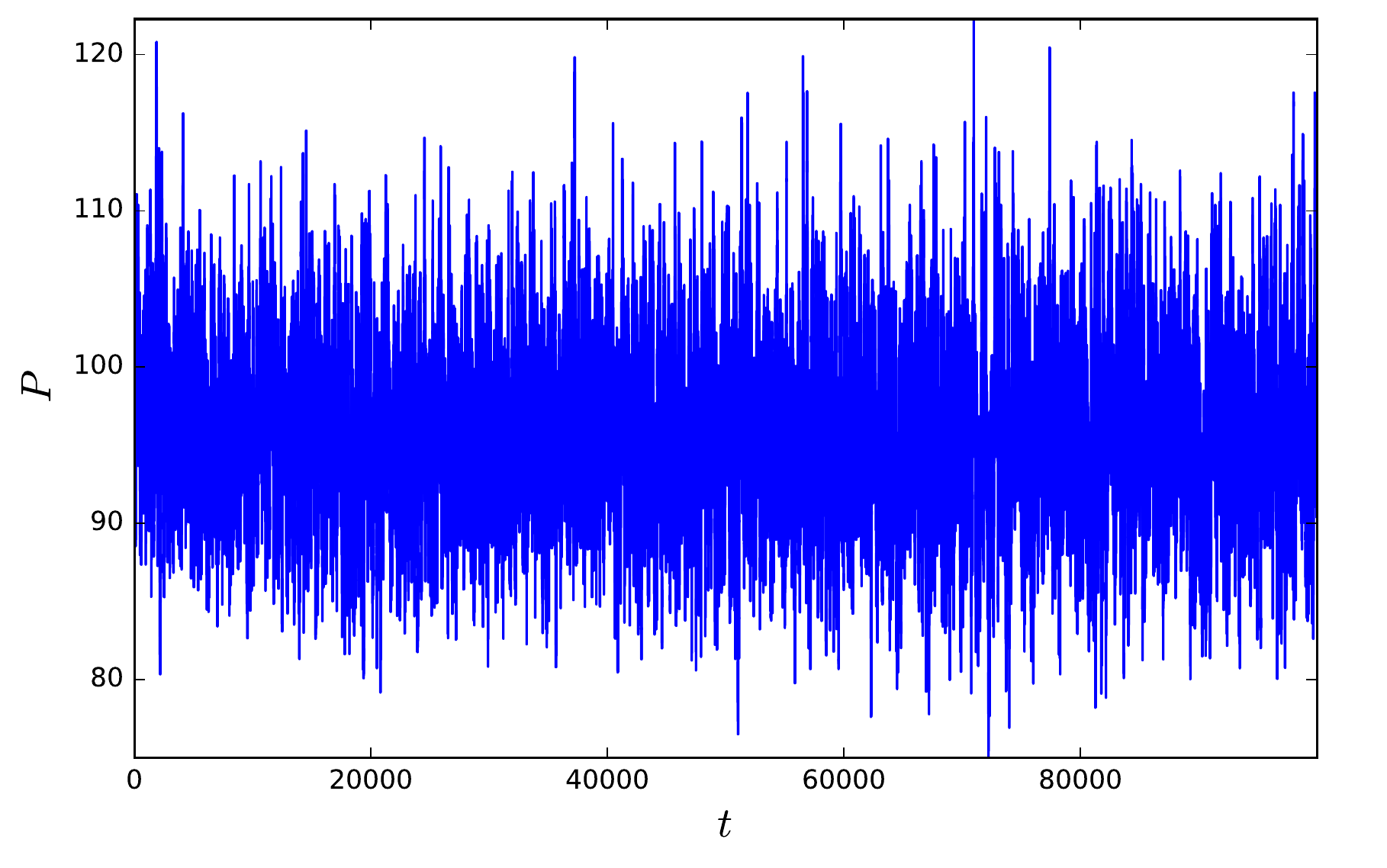}
    \includegraphics[width=8cm, height=5cm]{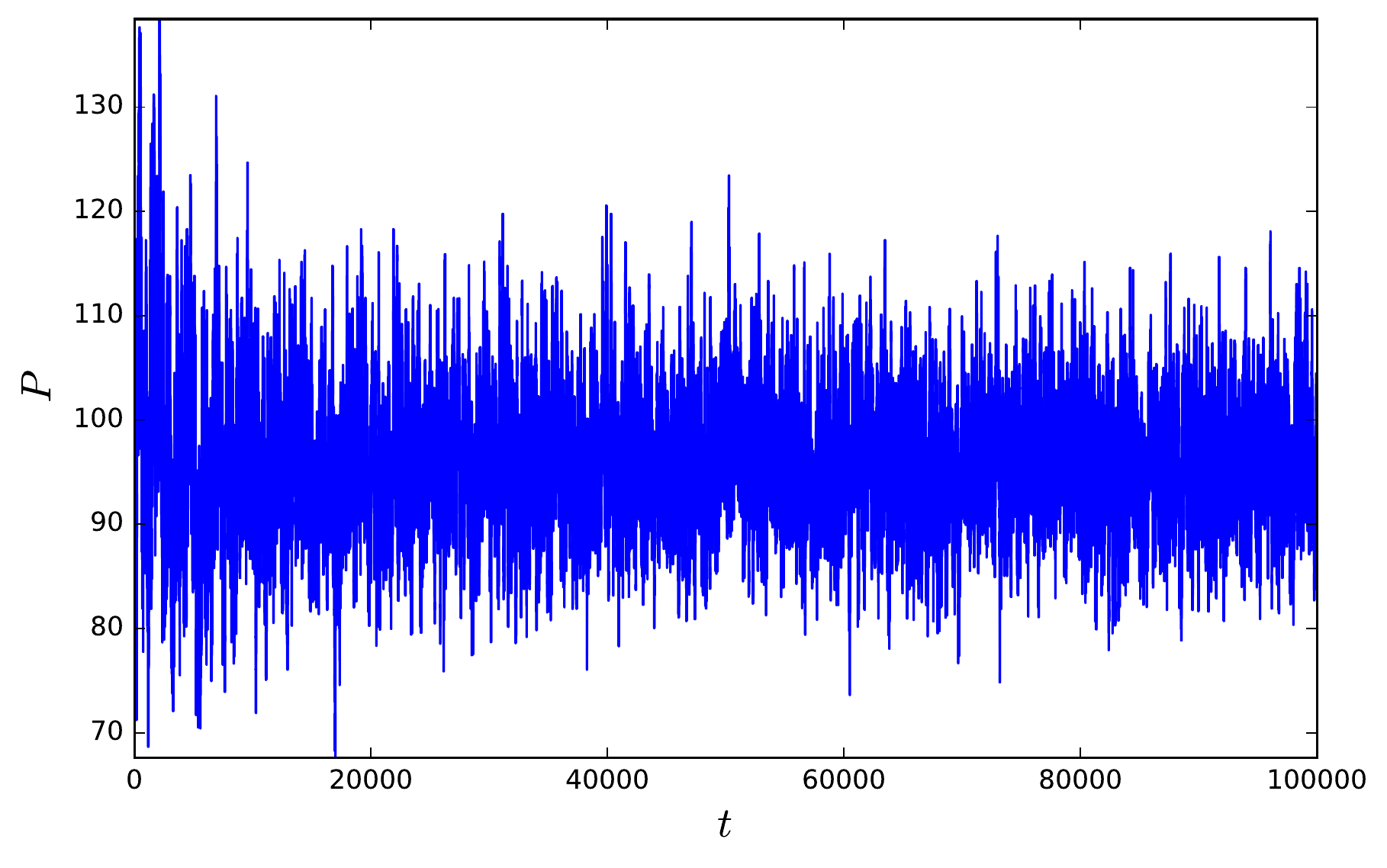}
 \caption{\label{stocks_section_4} Stock time series for the initial
 (top to bottom, left to right) identical, uniform, normal, and Pareto wealth allocations
 in section \ref{section_4}.
 The stock price exhibits a mean-reverting behavior around the equilibrium
 price $\sim 96$.
}
\end{figure}

\begin{figure}
 \includegraphics[width=8cm, height=8cm]{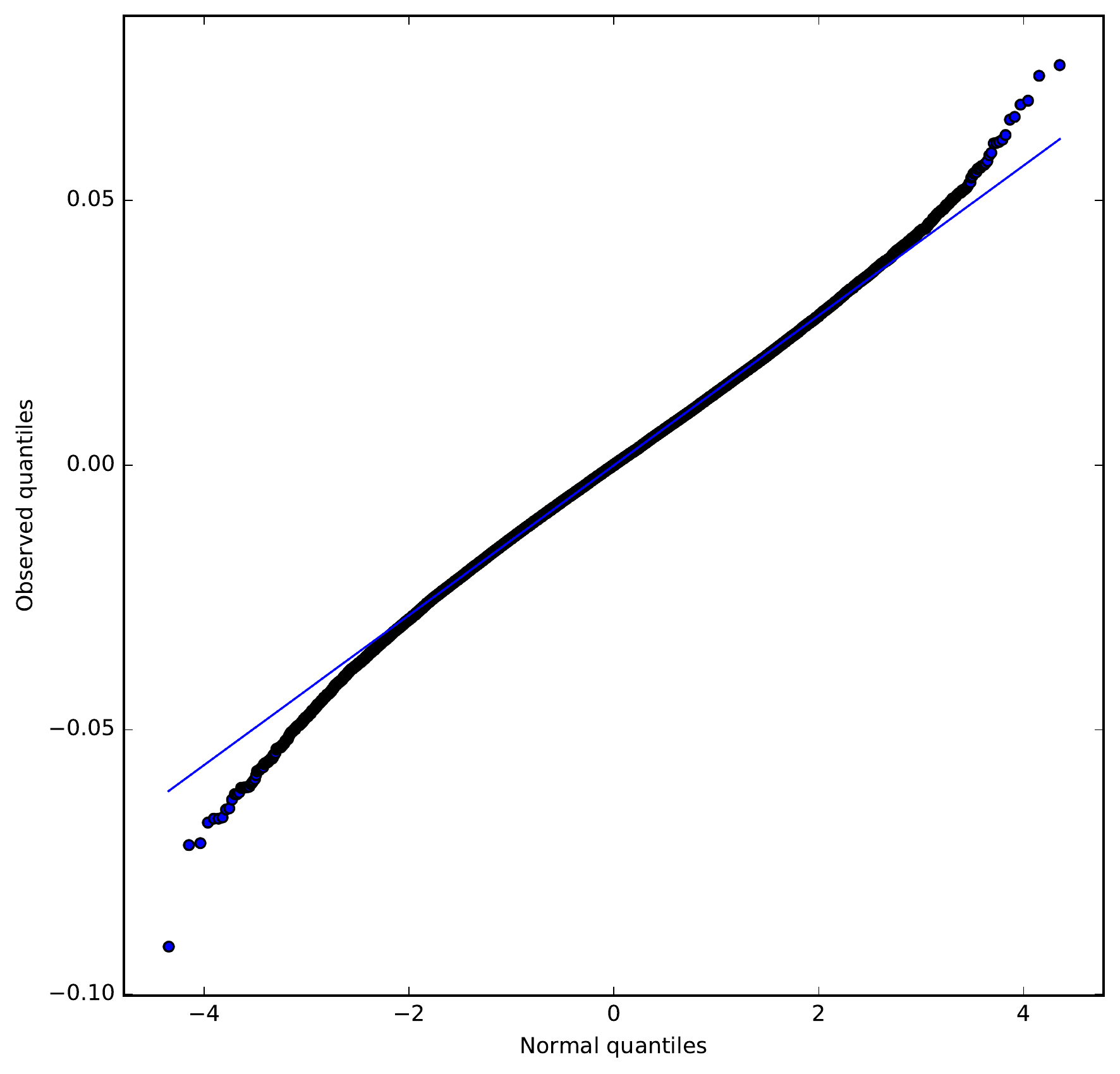}
  \includegraphics[width=8cm, height=8cm]{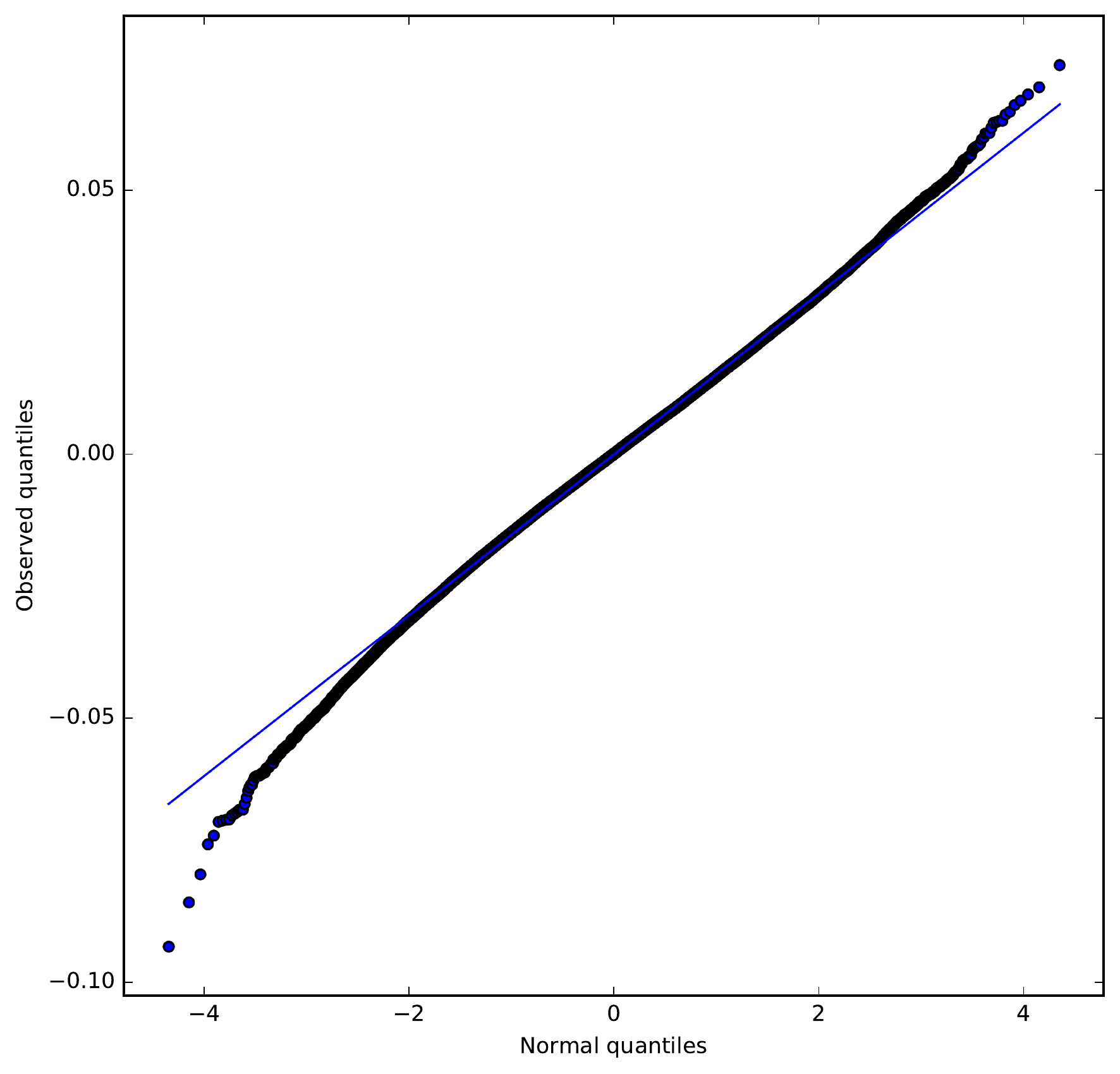}
   \includegraphics[width=8cm, height=8cm]{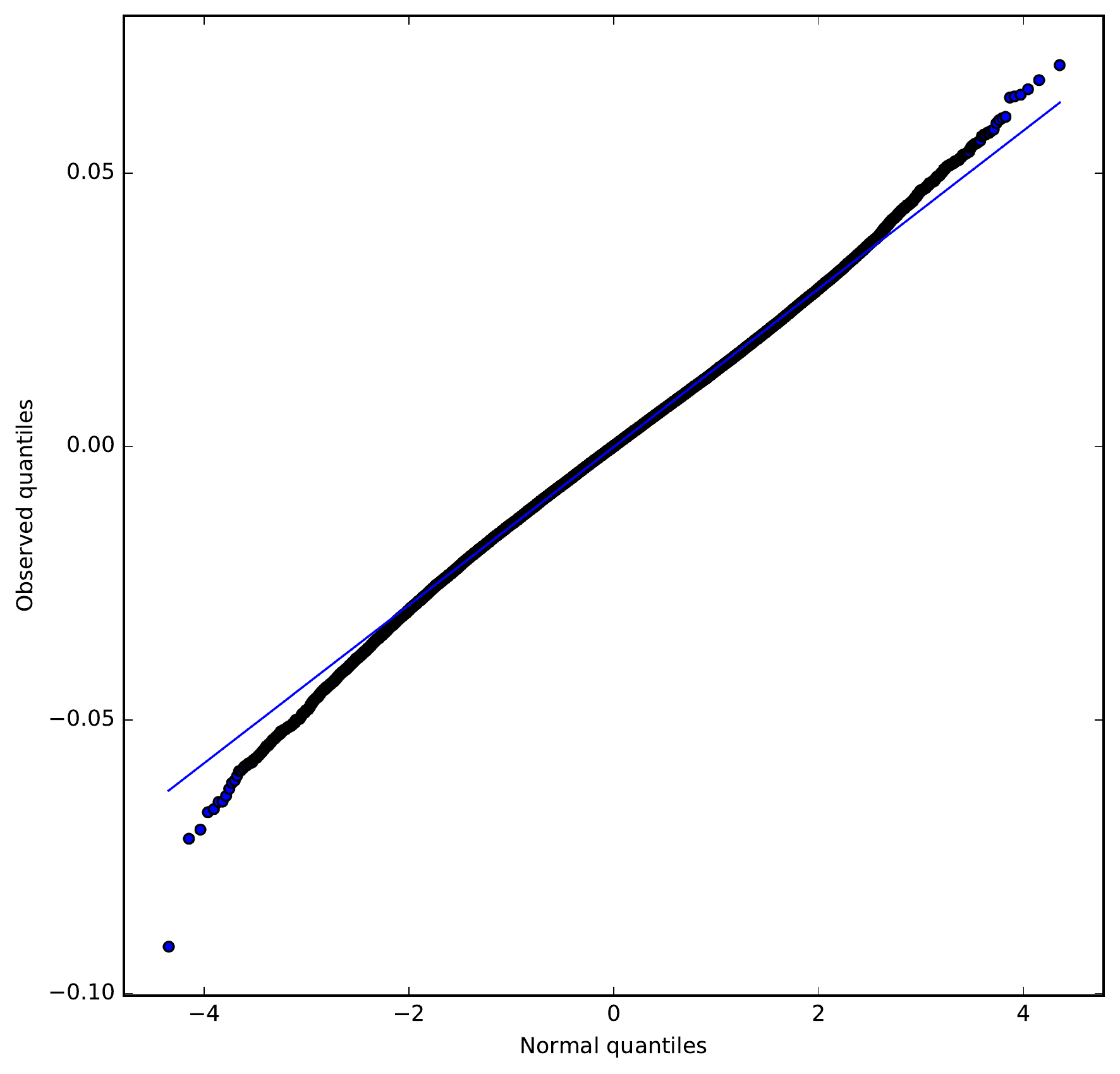}
    \includegraphics[width=8cm, height=8cm]{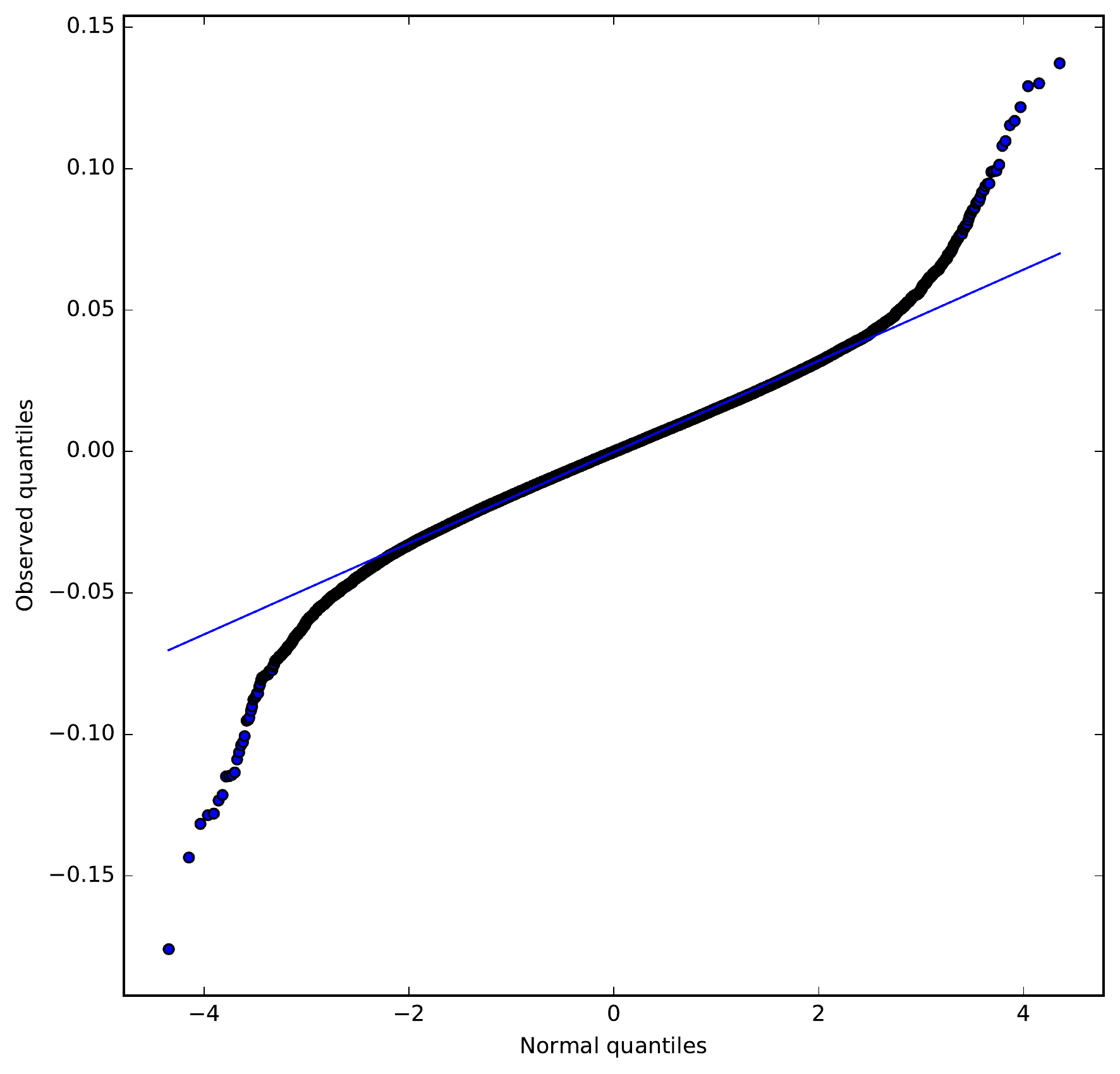}
 \caption{\label{QQ_section_4} Quantile-quantile plots for the initial
 (top to bottom, left to right) identical ($R^2=0.999$, sd=$0.0142$),
 uniform ($R^2=0.999$, sd=$0.0152$),
 normal ($R^2=0.999$, sd=$0.0145$),
 and Pareto ($R^2=0.992$, sd=$0.0165$) wealth allocations in section \ref{section_4}.}
\end{figure}

\begin{figure}
 \includegraphics[width=8cm, height=5cm]{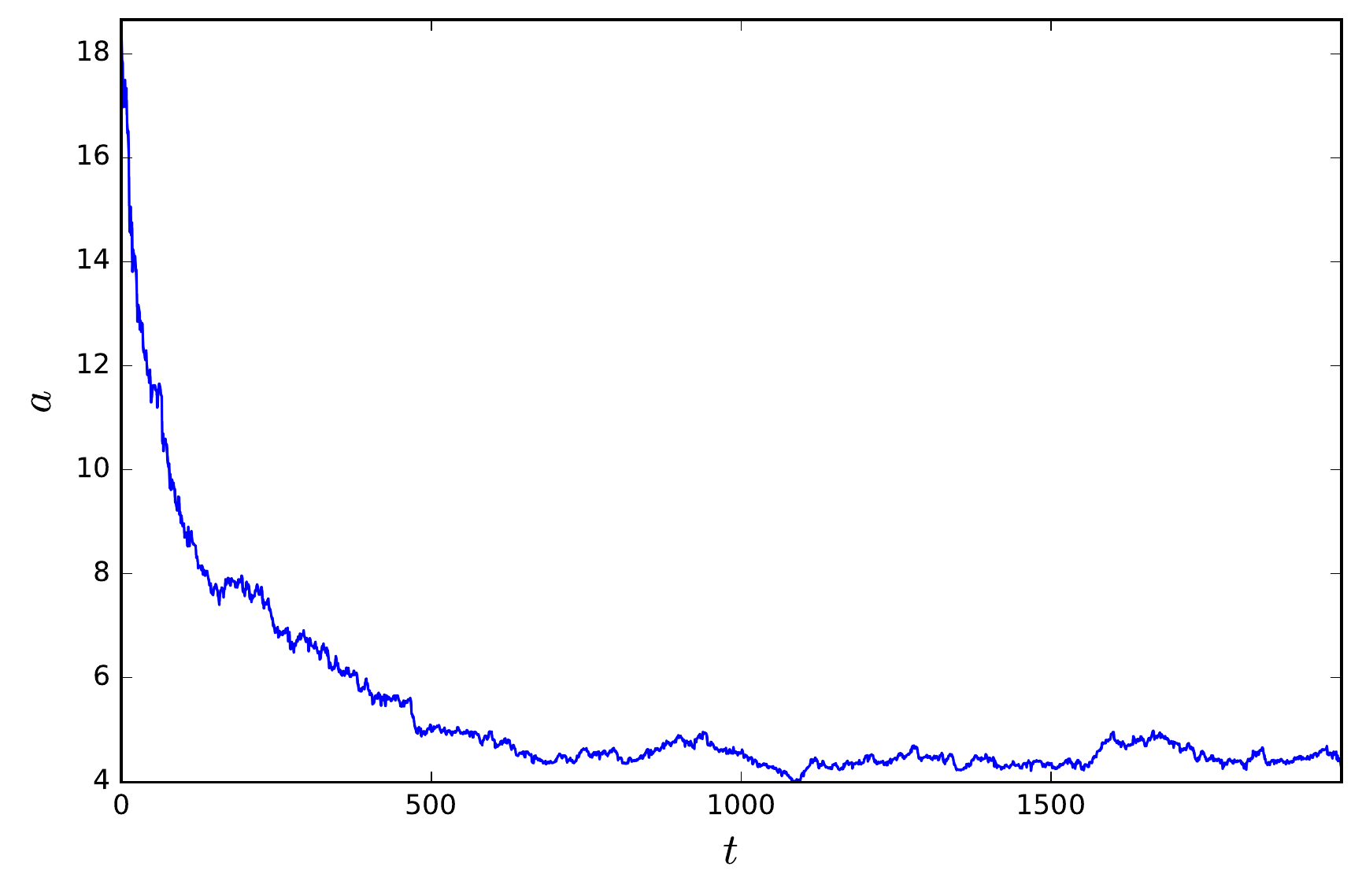}
  \includegraphics[width=8cm, height=5cm]{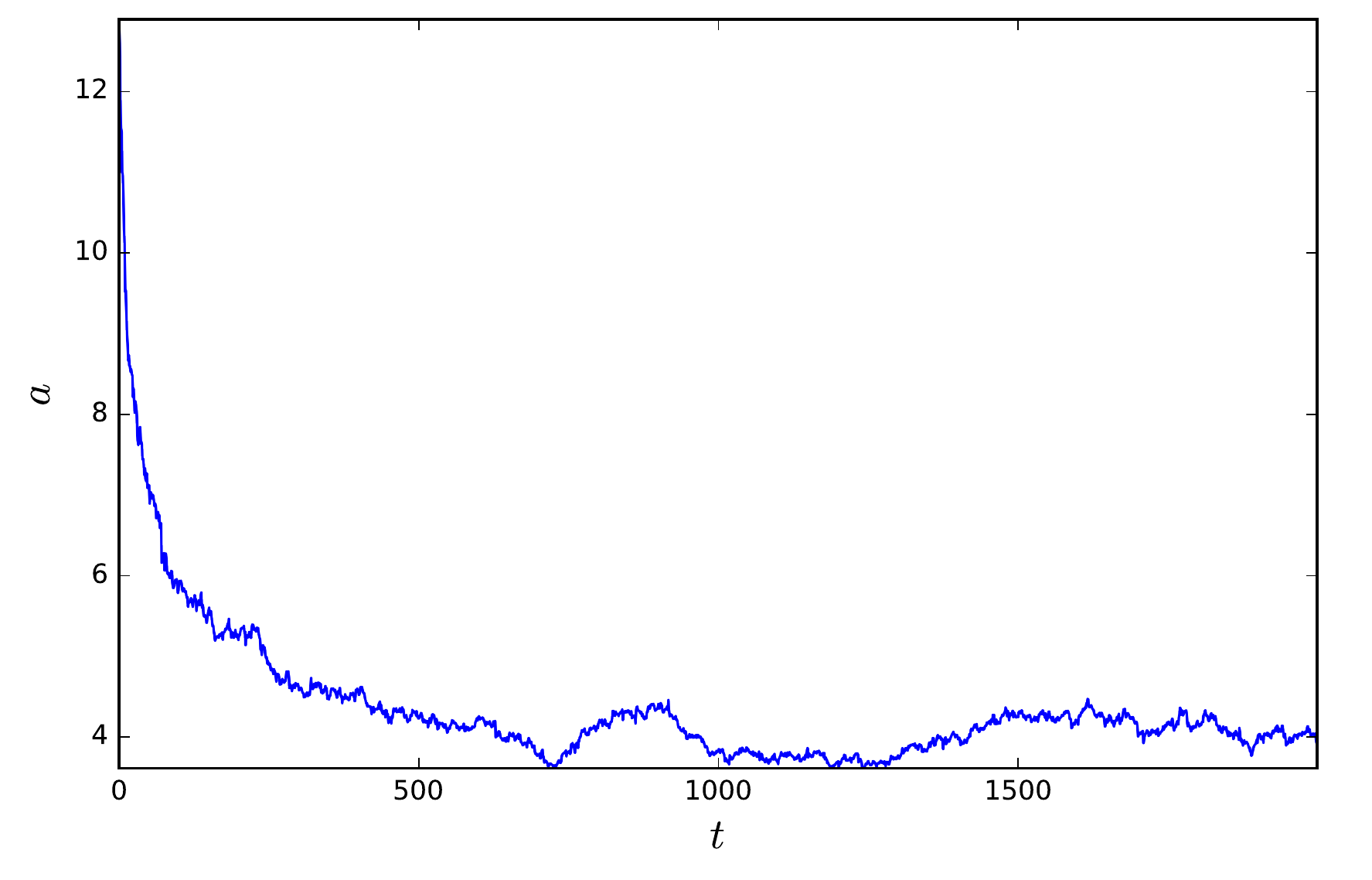}
   \includegraphics[width=8cm, height=5cm]{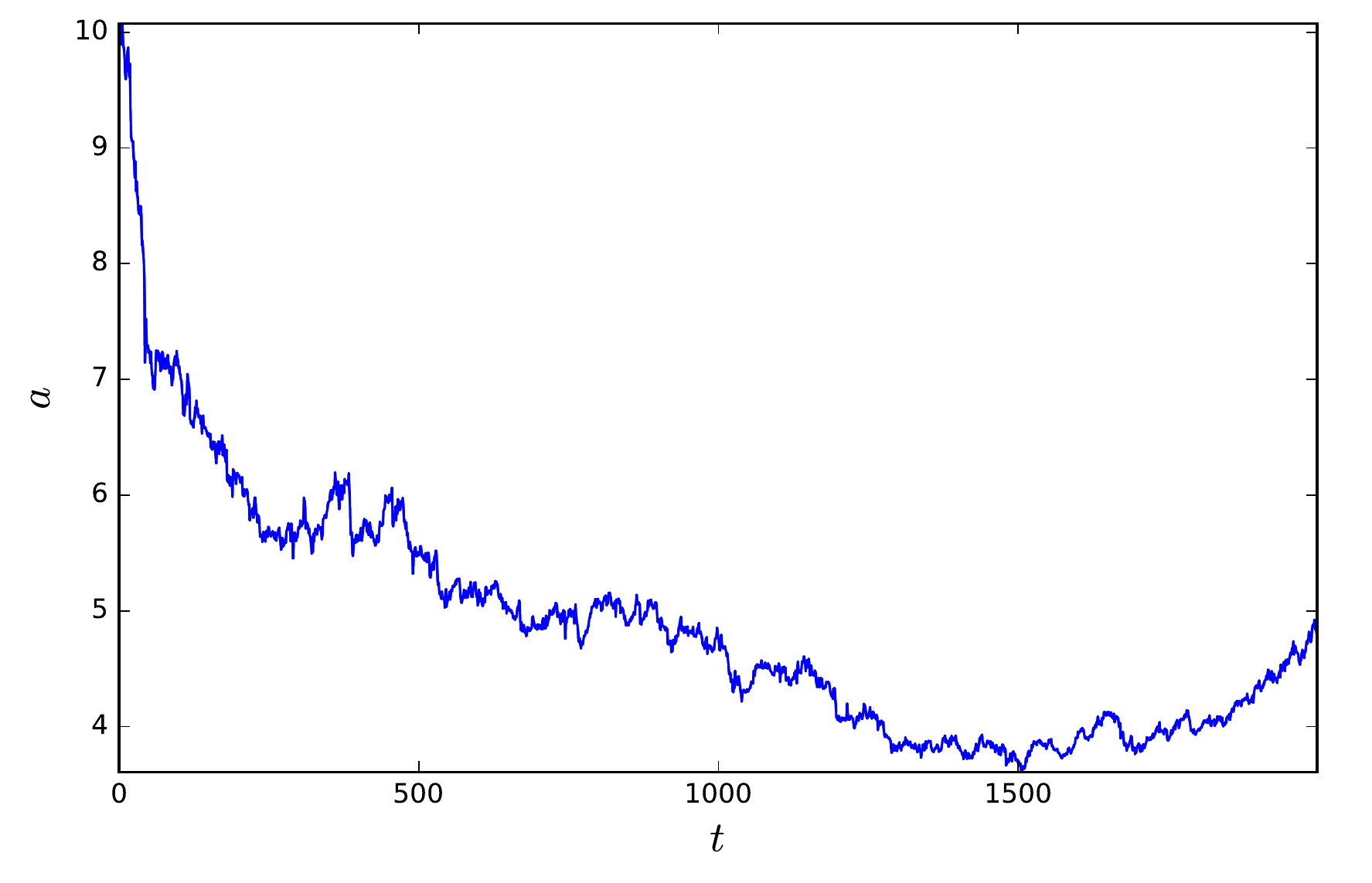}
  \includegraphics[width=8cm, height=5cm]{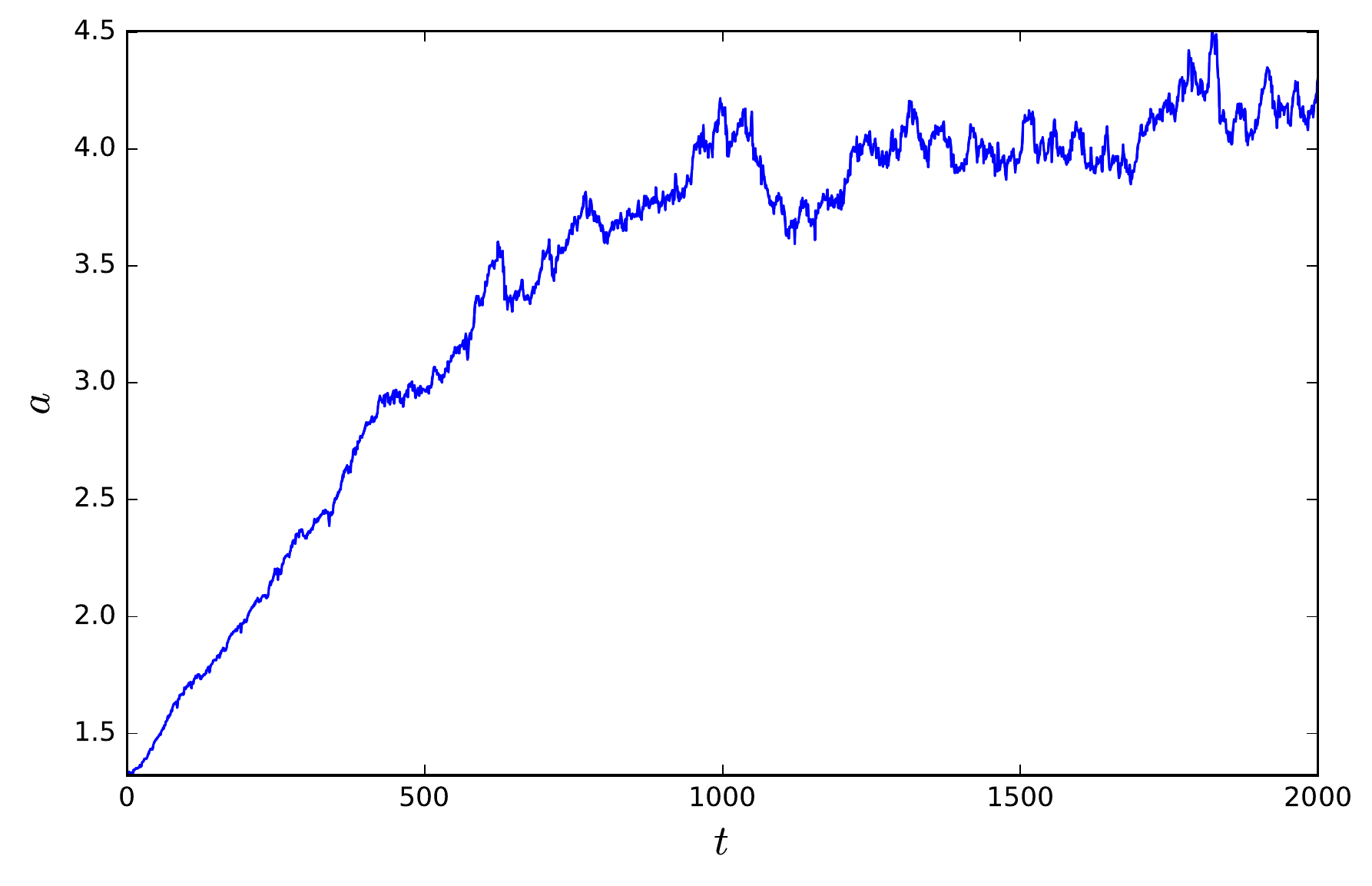}
 \caption{\label{Pareto_powers_section_4} Evolution of the Pareto
 tail (richest 25\% of the agents)
 fit power for the initial
 (top to bottom, left to right) identical (starting from step 30), uniform, normal, and Pareto wealth allocations
 in section \ref{section_4}. The measurements are taken every $50$ steps.}
\end{figure}

\begin{figure}
 \includegraphics[width=8cm, height=5cm]{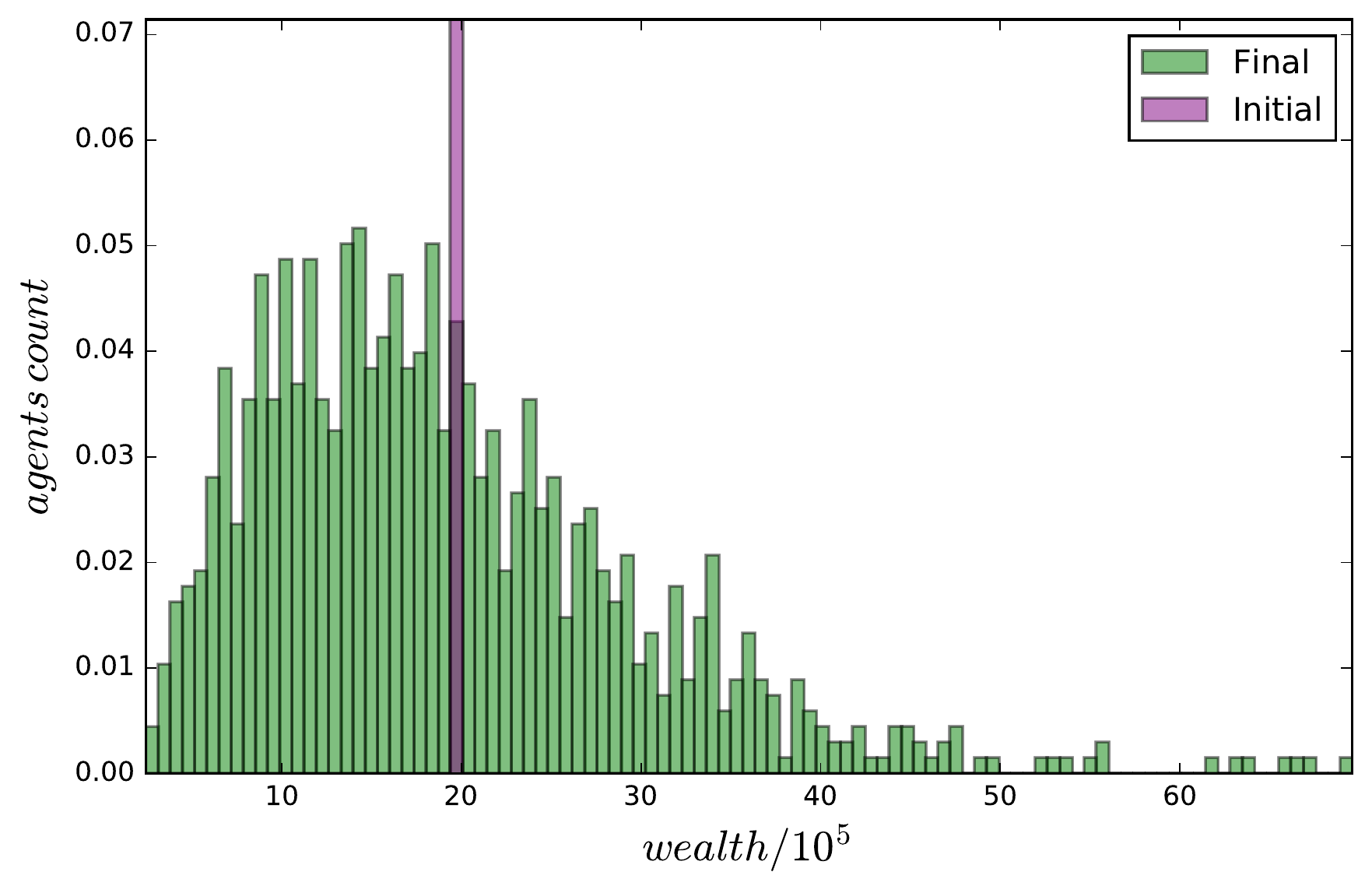}
  \includegraphics[width=8cm, height=5cm]{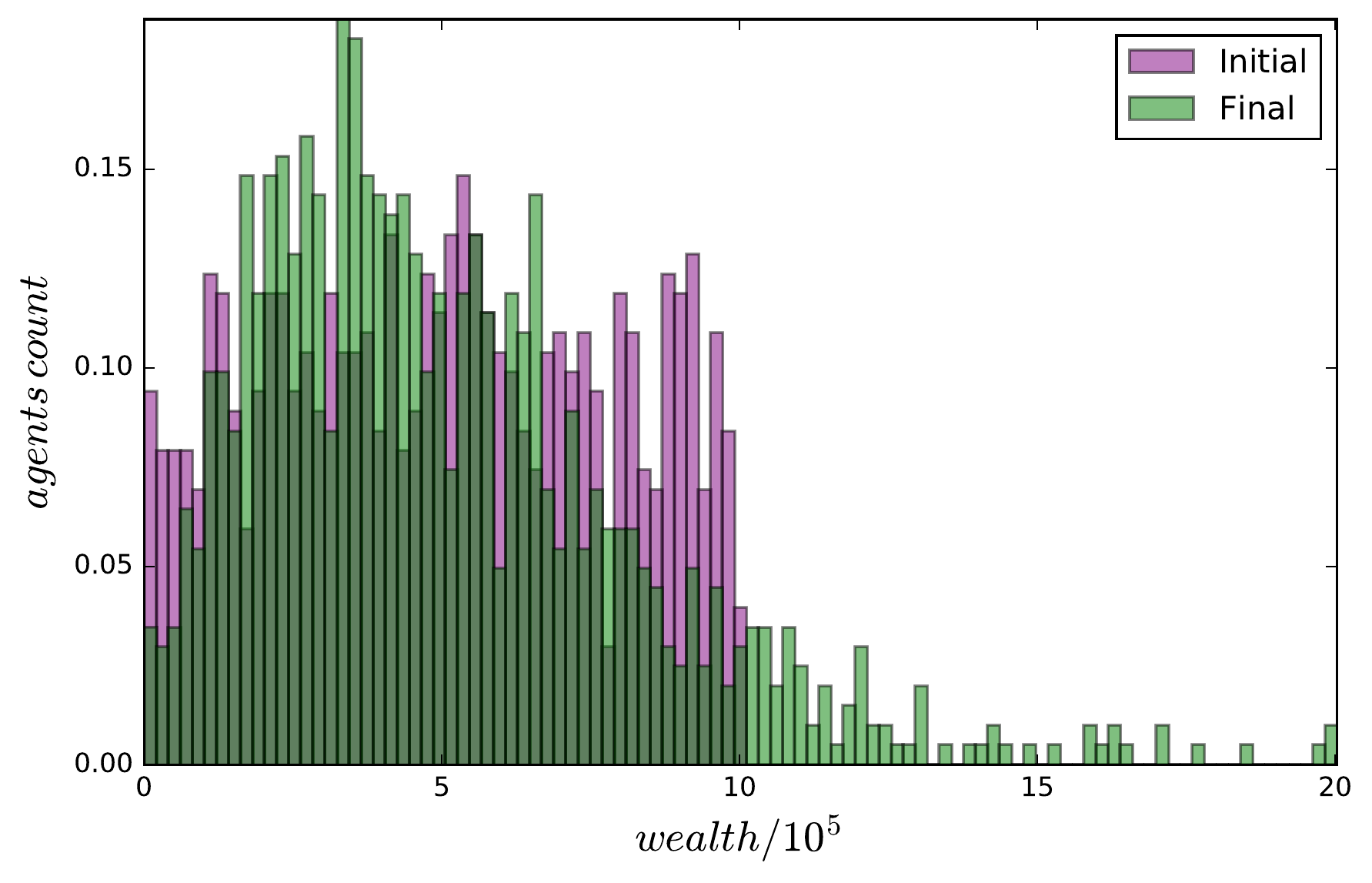}
   \includegraphics[width=8cm, height=5cm]{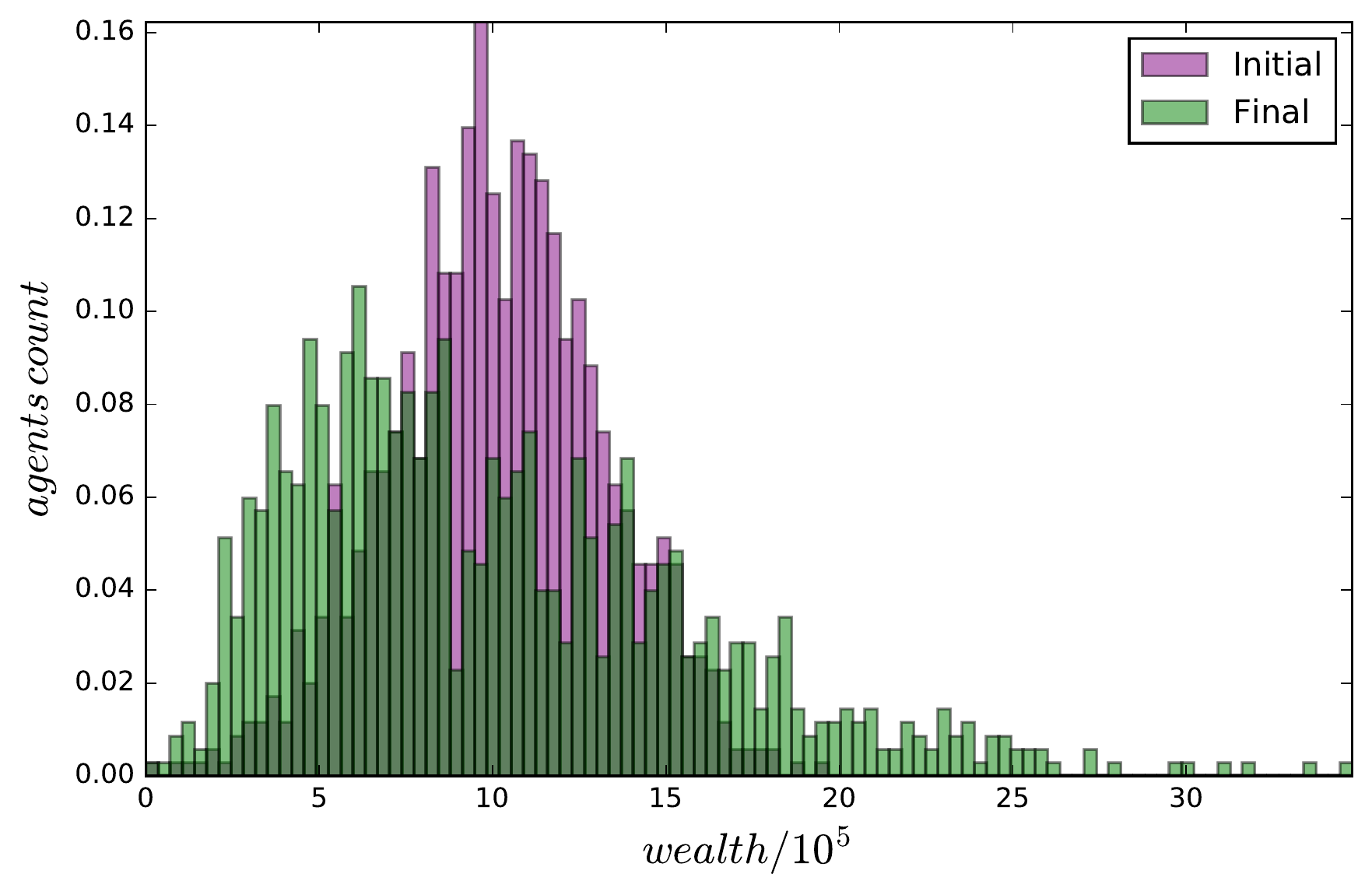}
    \includegraphics[width=8cm, height=5cm]{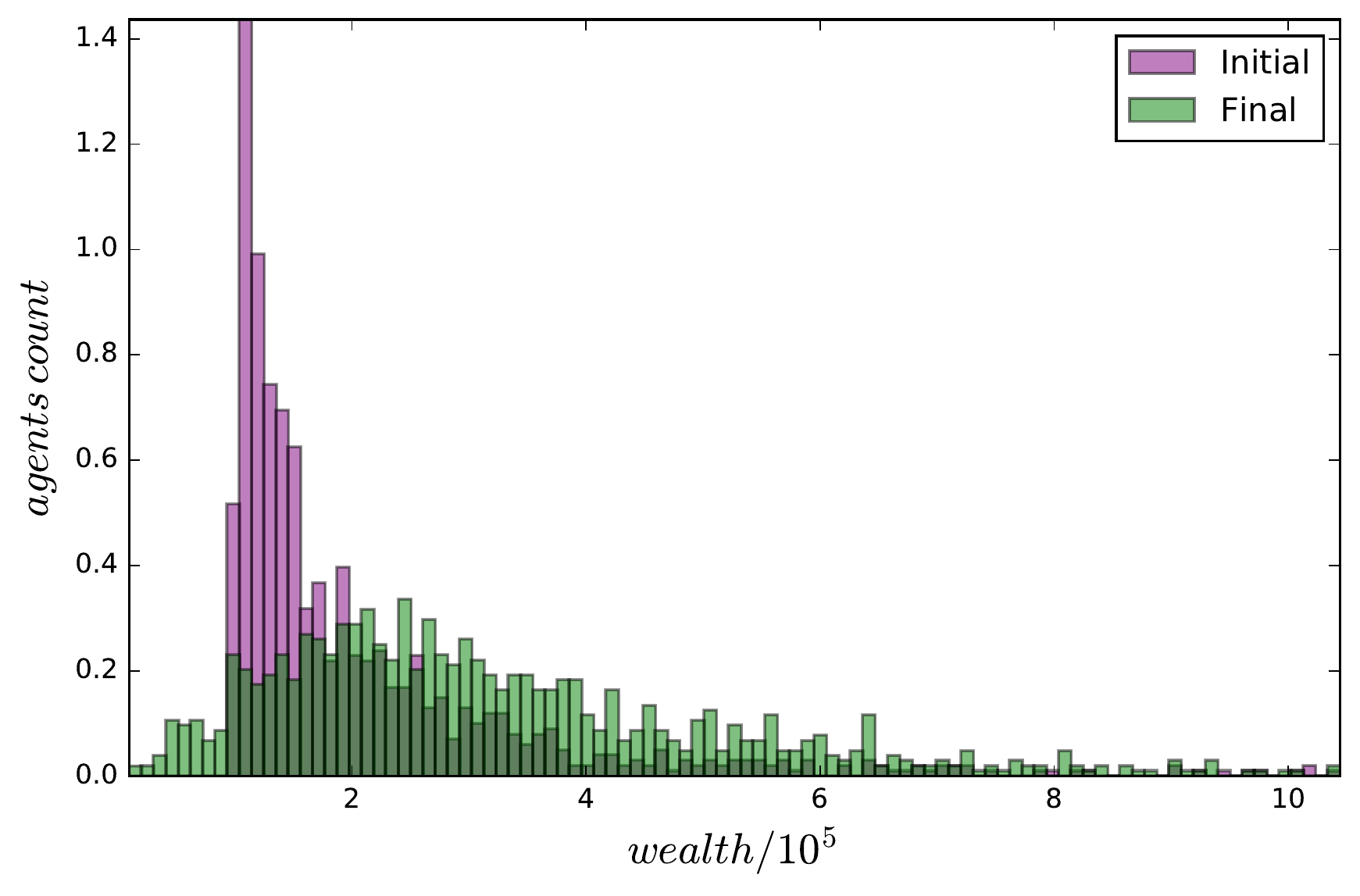}
 \caption{\label{Wealth_histogram_section_4} 
 Initial and final wealth histogram for the starting
 (top to bottom, left to right) identical, uniform, normal, and Pareto wealth allocations
 in section \ref{section_4}.}
\end{figure}

In this section we are going to focus on the wealth dynamics
in the participation sentiment-driven market environment set up in section \ref{Market_environment}.
We will study how the
market activity of $N=1000$ agents influences the
wealth distribution dynamics over $T=10^5$ simulation steps.
In each of the considered cases half of the initial portfolio value will be allocated in stock,
and half in cash.
We set the interest rate and the dividend yield identically to zero, $r=0$, $d=0$.
\footnote{Since in this section we are interested in the wealth dynamics due to the
trading activity, we need to have a model which will illustrate
reallocation of a given amount of resources in the system.
That is why we need to set $r=0$, $d=0$, so that the amount of cash
in the system stays constant.}

We will also
assume that the breaking news volatility is identical
to the calm volatility. Equivalently, we will assume
that the volatility process (\ref{sigma_process}) has infinite arrival time, $\lambda=\infty$.
We set the calm volatility to be
\begin{equation}
\sigma_c= {\cal N}(0.05,0.001)\,.\label{calm_volatility_4}
\end{equation}
The inverse mean arrival time for the agents's market participation sentiment process (\ref{rho_process})
will be set to
\begin{equation}
\rho^{-1}={\cal N}(0.1,0.01)\,.\label{poisson_agent_participation_process}
\end{equation}

In this section we will set the buy/sell dis-balance
sentiment (\ref{buy_sell_disbalance}) to be identically zero, $\psi=0$.
Then from (\ref{equilibrium_buy_sell}) it follows that the
equilibrium stock price around which the stock time series is expected
to exhibit a mean-reverting behavior is equal to $M/S$,
where $M$ is the total amount of cash and $S$ is the total
number of shares outstanding.
Since $r=0$ and $d=0$ the total amount
of cash $M$ is time-independent. The total number
of shares outstanding $S$ is also a constant.
Initially every agent receives half a portfolio in cash, and half in stock,
and therefore we expect the stock
price to exhibit a mean-reverting behavior around the equilibrium value $M/S=P_0$, equal to the initial stock price.

However, running the simulation we observe that the stock price
exhibits mean-reverting behavior around the equilibrium value $\sim 96$,
see fig. \ref{stocks_section_4}.
This is an artifact of the limit price choice, allocating
a higher probability to the price lower than
the last stock price, as discussed below eq. (\ref{limit_price_submitted}).

Besides the stock price times series we also are interested in
the stock returns distribution and the wealth dynamics.
We study the stock returns distribution by plotting
a quantile-quantile plot against the standard normal quantiles
for the logarithmic stock returns, see fig. \ref{QQ_section_4}. 
We observe that the stock returns are well approximated by the log-normal
distribution, with the QQ fitting line having a zero intercept (mean
return), and the slope (standard deviation) around $\sigma_r=0.015$. \footnote{Notice that the limit
price uncertainty (\ref{calm_volatility_4}) is related
to stock volatility $\sigma_r$ as $\sigma_c\simeq 3\sigma_r$.
This relation can be compared with \cite{Raberto2000},
where the agents were measuring the past stock price volatility and
estimating their price uncertainty perception $\sigma_c=k\sigma_r$,
with the input choice $k=3.5$. In our model this value of $k$ is emergent
rather than put in by hand.} We will see below that turning on the jump
diffusion sentiment will introduce significant deviation of the stock
log-returns from the normal distribution.

The most prominent feature of fig. \ref{QQ_section_4} is that the stock returns
are most deviating from the log-normal distribution for the case when the initial
wealth of the agents is Pareto-distributed. This property of the (small-exponent)
Pareto society will be confirmed in other simulations in the subsequent sections.
It is consistent with the observation \cite{RabertoCF2003} that since the Pareto
society is realistically expected to have a few very rich agents, when those agents
will be chosen to trade they will submit a large-sized orders. Such an orders
will skew the supply-demand intersection point, resulting in a large price change.

We notice that regardless of the considered choices of the initial wealth allocation,
the wealth distribution quickly acquires a Pareto tail with the exponent $a$
undergoing dynamics demonstrated in fig. \ref{Pareto_powers_section_4}.
Eventually in all of the considered cases the Pareto tail exponent approaches the value $a\simeq 4$.
This means that the exponent will go down (thickening the wealthy tail)
for the considered initial identical, uniform, and normal wealth distribution.
For the initial $a=1.5$ Pareto distribution it will go up, thinning the wealthy tail. 
We plot the initial and final wealth allocation histograms in fig. \ref{Wealth_histogram_section_4}.

\section{Including the buy/sell sentiment in the jump volatility market}\label{section_5}

\begin{figure}
 \includegraphics[width=8cm, height=5cm]{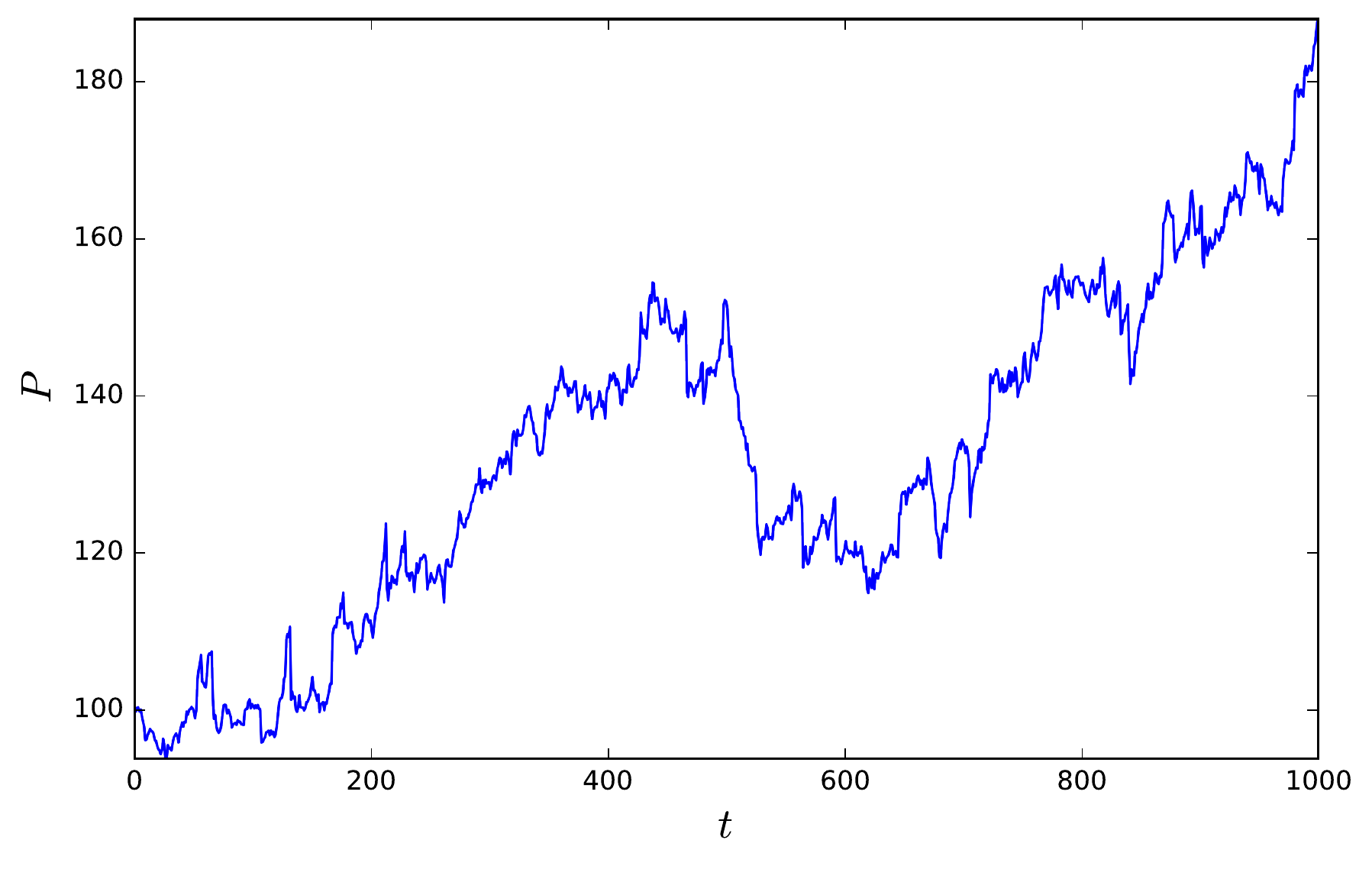}
  \includegraphics[width=8cm, height=5cm]{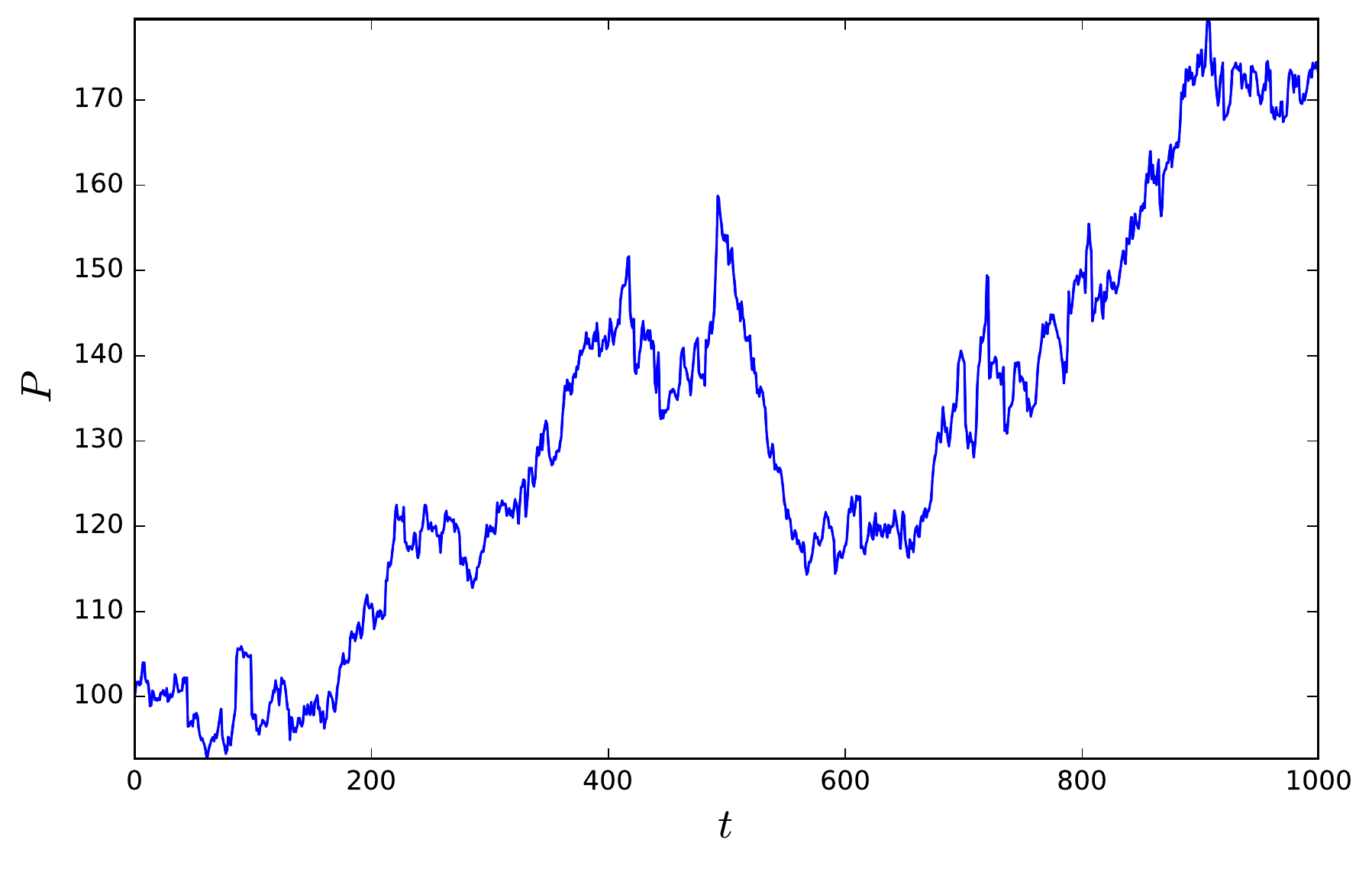}
   \includegraphics[width=8cm, height=5cm]{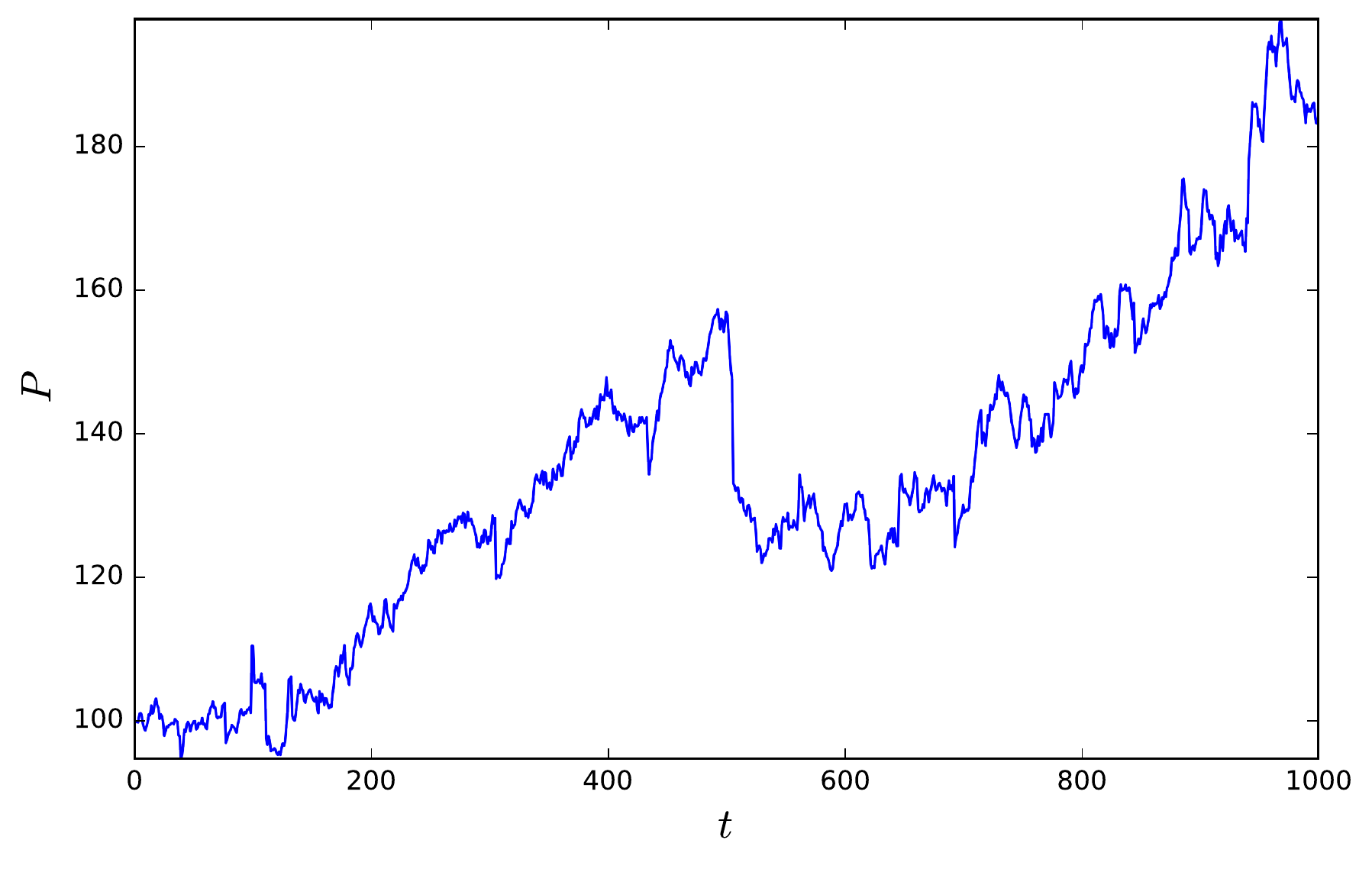}
    \includegraphics[width=8cm, height=5cm]{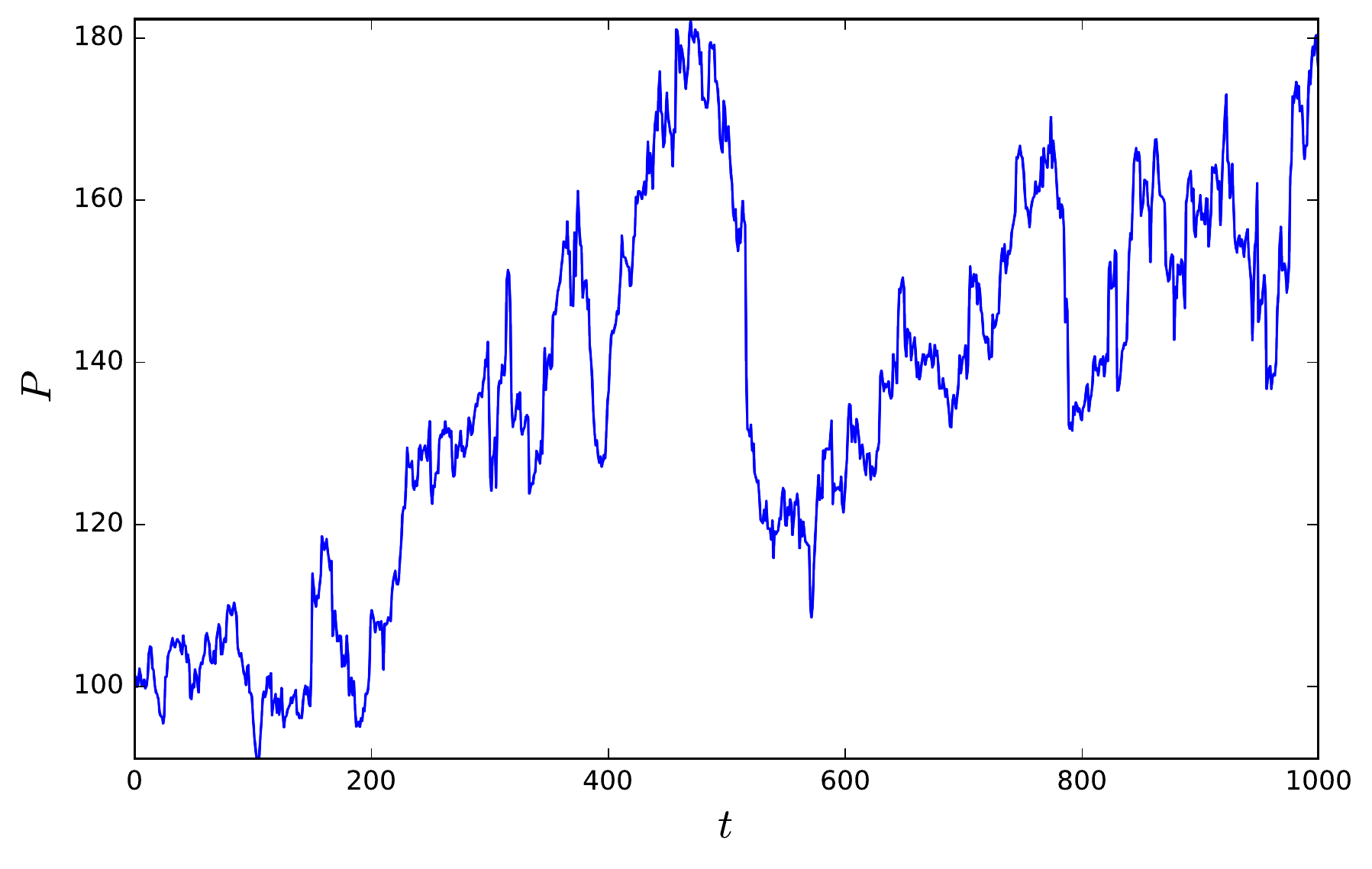}
 \caption{\label{stocks_section_5} Stock time series for the initial
 (top to bottom, left to right) identical, uniform, normal, and Pareto wealth allocations
 in section \ref{section_5}.
}
\end{figure}

\begin{figure}
 \includegraphics[width=8cm, height=8cm]{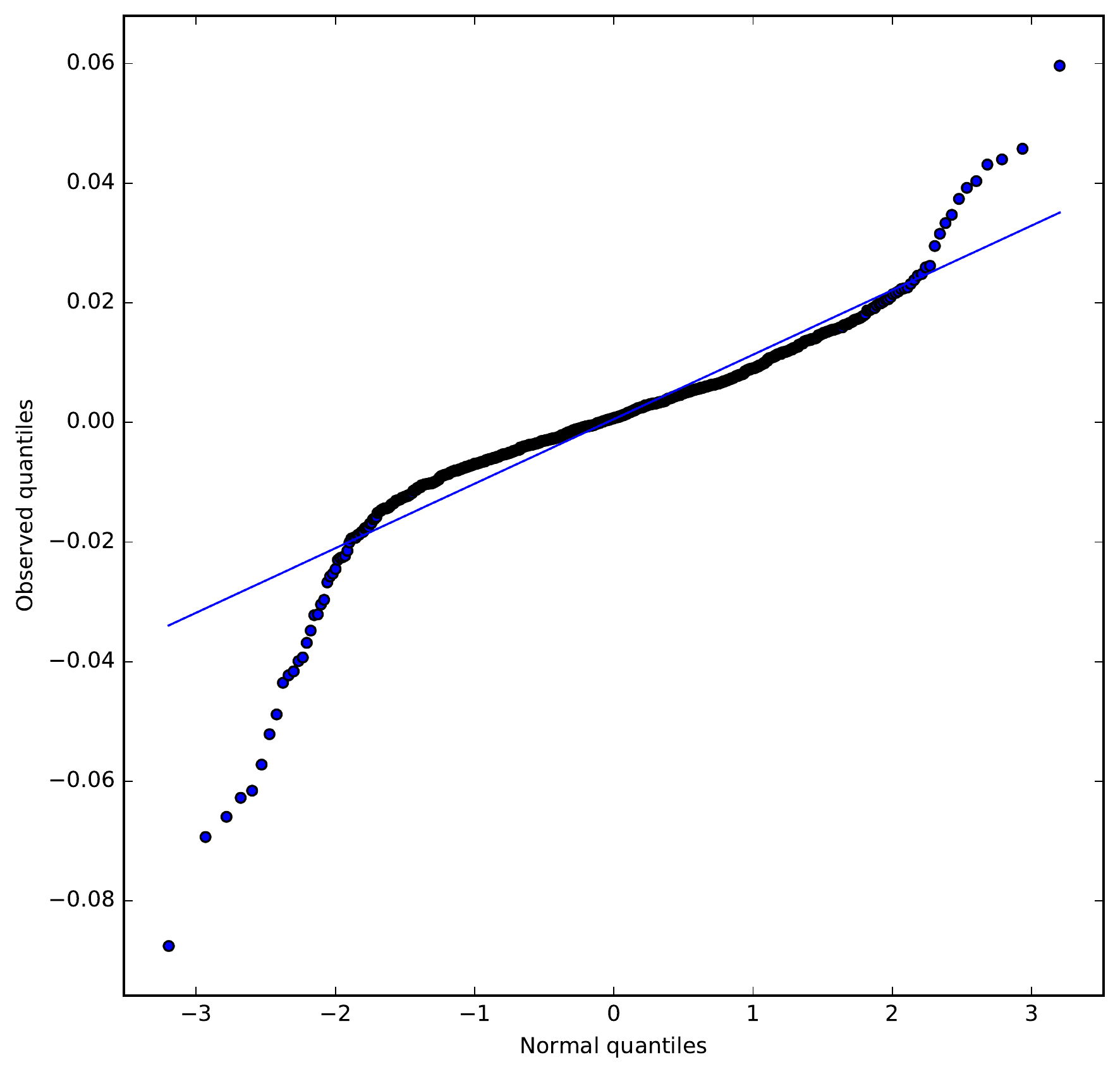}
  \includegraphics[width=8cm, height=8cm]{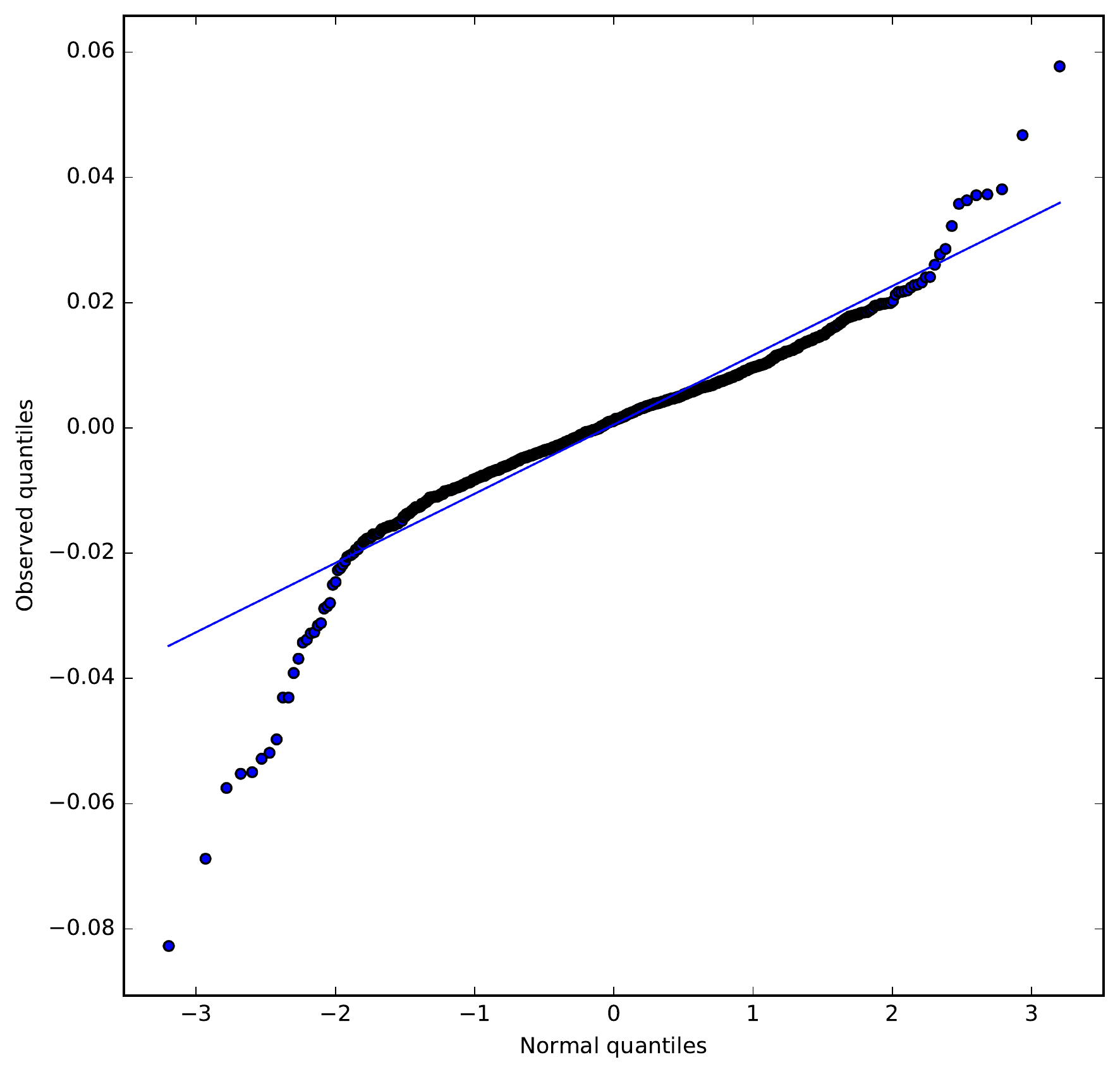}
   \includegraphics[width=8cm, height=8cm]{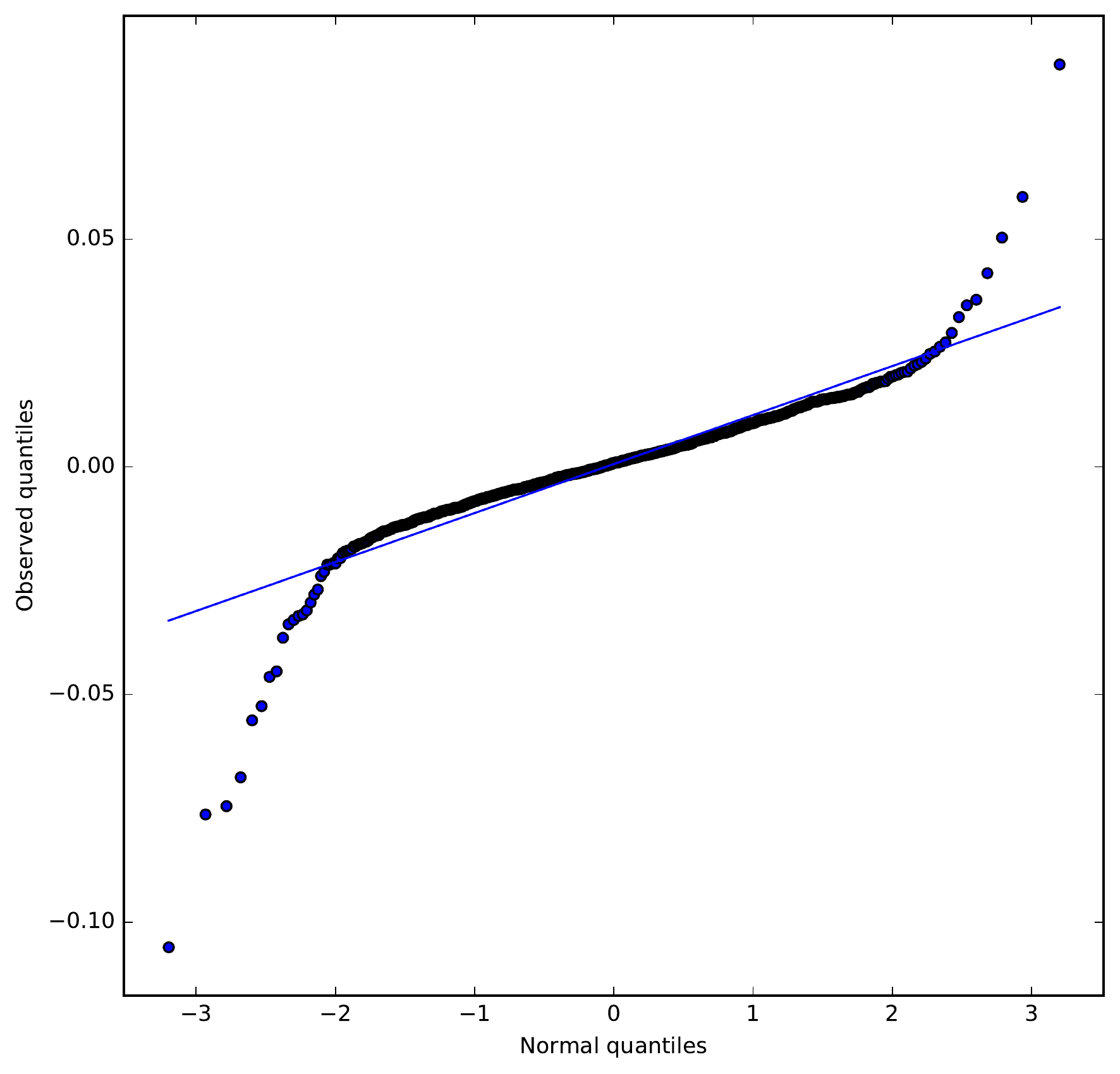}
    \includegraphics[width=8cm, height=8cm]{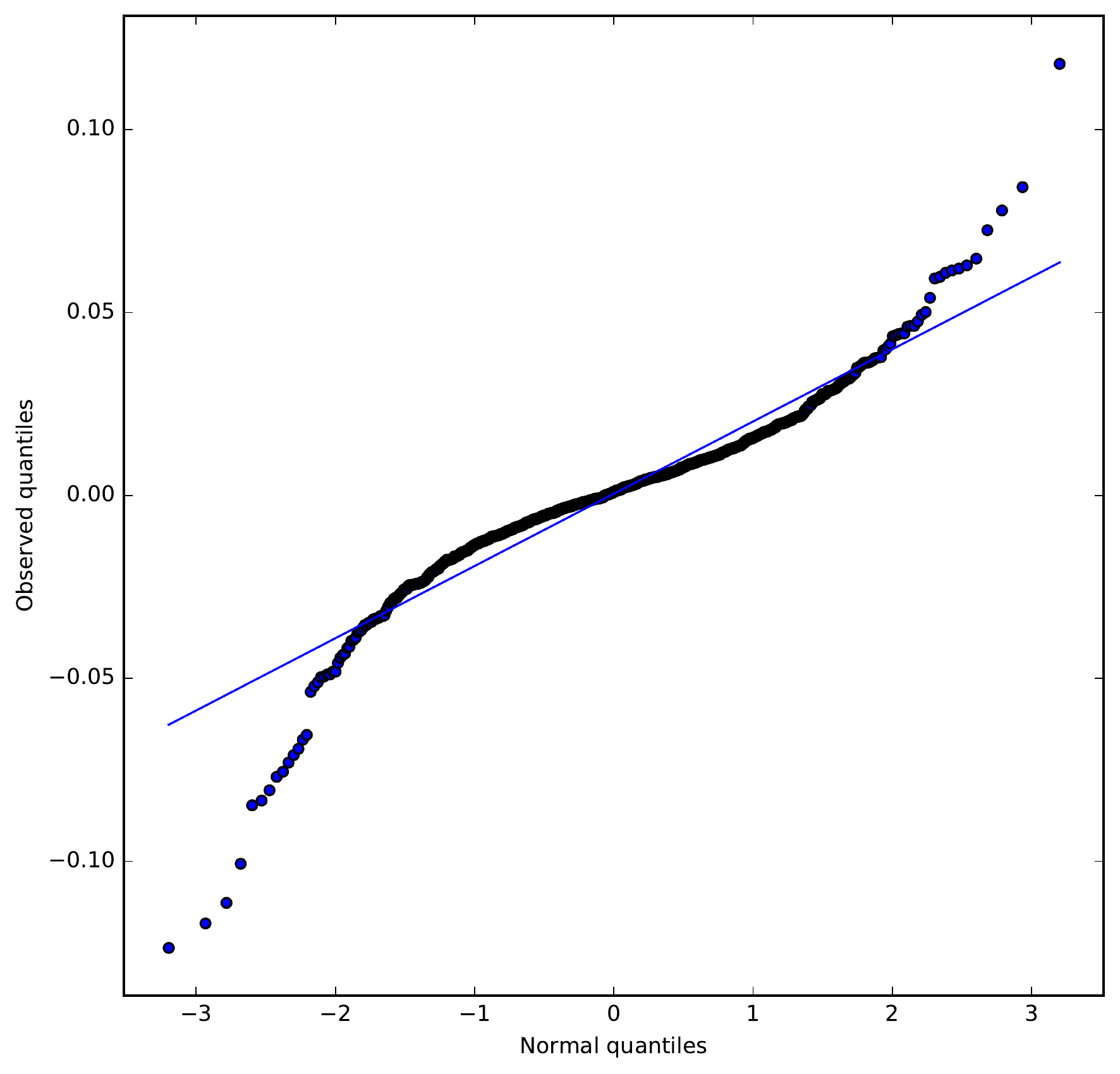}
 \caption{\label{QQ_section_5} Quantile-quantile plots for the initial
 (top to bottom, left to right) identical ($R^2=0.93$, mean=$-0.0006$, sd=$0.01$),
 uniform ($R^2=0.95$, mean=$-0.0005$, sd=$0.01$),
 normal ($R^2=0.92$, mean=$-0.0006$, sd=$0.01$),
 and Pareto ($R^2=0.95$, mean=$-0.0006$, sd=$0.02$)
  wealth allocations in section \ref{section_5}.}
\end{figure}

\begin{figure}
 \includegraphics[width=8cm, height=8cm]{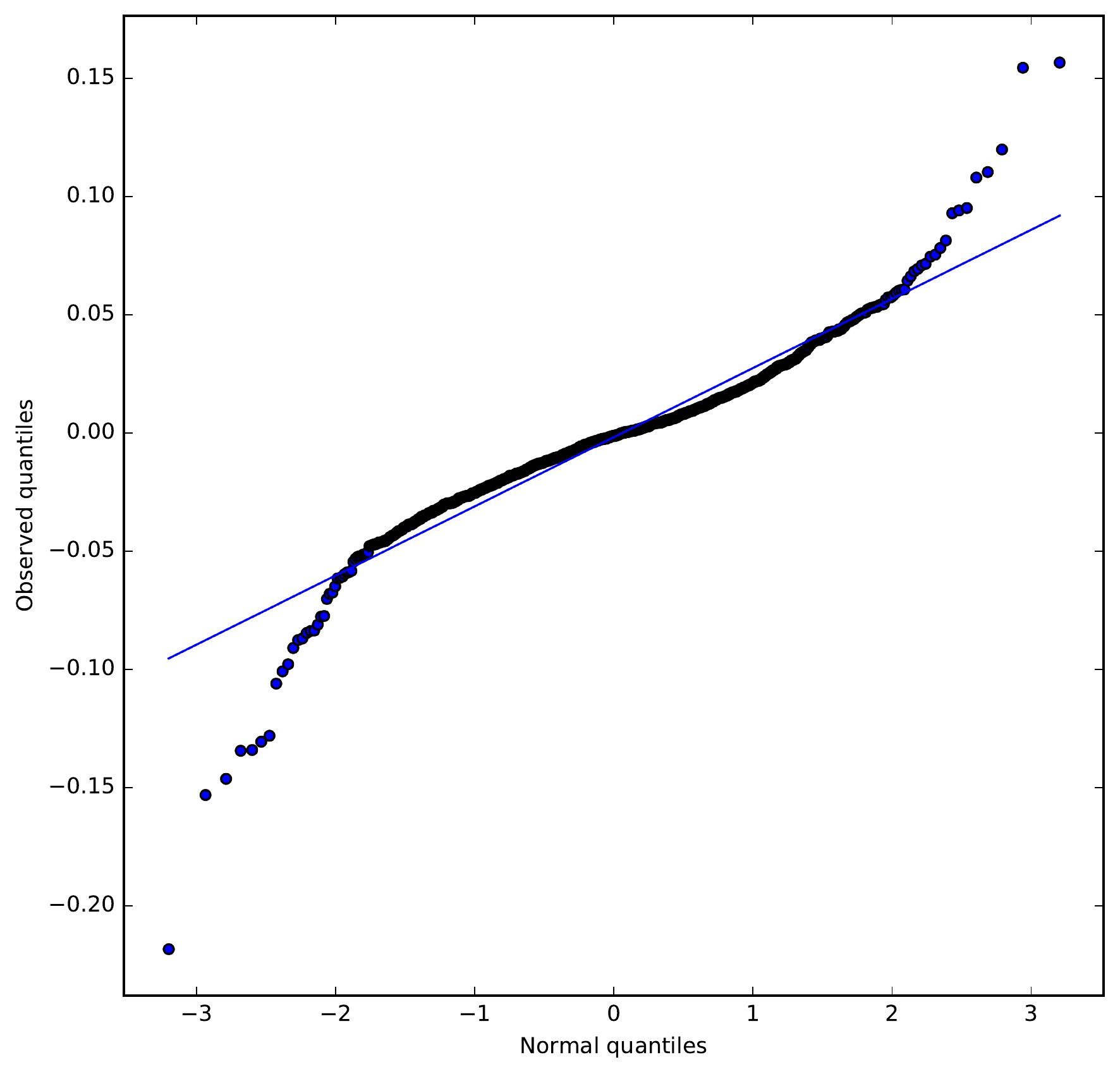}
  \includegraphics[width=8cm, height=8cm]{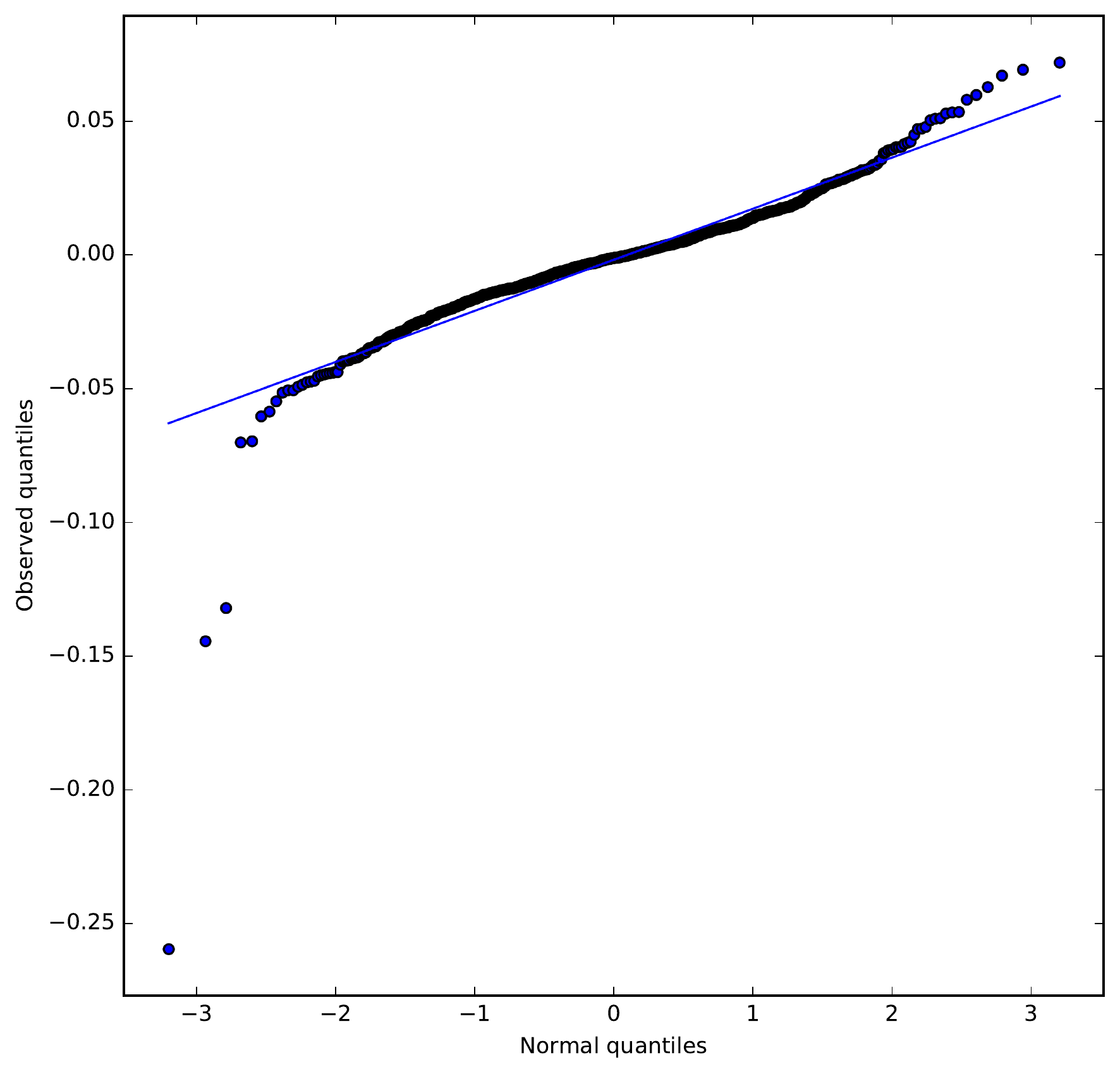}
 \caption{\label{QQ_TSLA} Quantile-quantile plots for: {\bf left} TSLA
 ($R^2=0.96$, mean=$0.0018$, sd=$0.029$),
 {\bf right} FB ($R^2=0.92$, mean=$0.0017$, sd=$0.019$),
 daily log returns over the period 
May 1, 2013-May 1, 2017.}
\end{figure}

In this section we are going to illustrate the influence
of the buy/sell sentiment and the breaking news volatility
sentiment on the stock price dynamics. We will consider
evolution of the system of  $N=1000$ agents
over $T=1000$ steps.
We will assume the per-step interest rate $r=10^{-3}$ and the vanishing dividend yield $d=0$.

We will also assume that the buy/sell sentiment is
  \begin{equation}
  \psi(t) = \left \{
  \begin{aligned}
    &{\cal N}(0,0.01), && \text{if}\ t\in [0,T/2] \\
    &{\cal N}(-0.3,0.01), && \text{if}\ t \in [T/2,T]
  \end{aligned} \right.
\end{equation} 
This illustrates a simple
attitude change in which at $t=T/2$ the stock receives a slight downgrade,
which establishes the sell sentiment $e^\psi=p_b/p_s\simeq 0.74$. We know that according to
(\ref{equilibrium_psi_price}) the change in the buy/sell
sentiment $\psi$ will result in the change of the expected
equilibrium stock price $P_e$. This effect is imposed on top
of the overall stock price up-trend due to the cash inflow coming
from the interest payments. 

The calm volatility will be $\sigma_c={\cal N}(0.05,0.001)$,
the breaking news volatility will be $\sigma_b={\cal N}(0.2,0.001)$.
The breaking news inverse mean arrival time will be $\lambda^{-1}={\cal N}(0.08,10^{-4})$.
The agents's trading inverse mean arrival time will be $\rho^{-1}={\cal N}(0.3,0.01)$.

In fig. \ref{stocks_section_5} we plot the simulated stock
price time series.
As expected, after step $t=500$ the stock price quickly declines to about $e^{-0.3}\simeq 0.74$
of its preceding value, in accord with (\ref{equilibrium_buy_sell}). Notice that
since all of the market participants experience the same sell over buy sentiment,
the stock decline is rather steep.  Also notice that in agreement with fig. \ref{QQ_section_4}
the Pareto society is the most volatile, due to the large orders submitted by exceptionally
rich agents of the Pareto society \cite{RabertoCF2003}.

In fig. \ref{QQ_section_5} we plot
the normal quantile-quantile plot of the stock log returns. 
As a frame of reference we provide the normal quantile-quantile plot
for the TSLA  and FB stocks daily returns 
over the period 
May 1, 2013-May 1, 2017, see fig. \ref{QQ_TSLA}.

\section{Subgroups of agents with different sentiments}\label{section_6}

\begin{figure}
 \includegraphics[width=8cm, height=5cm]{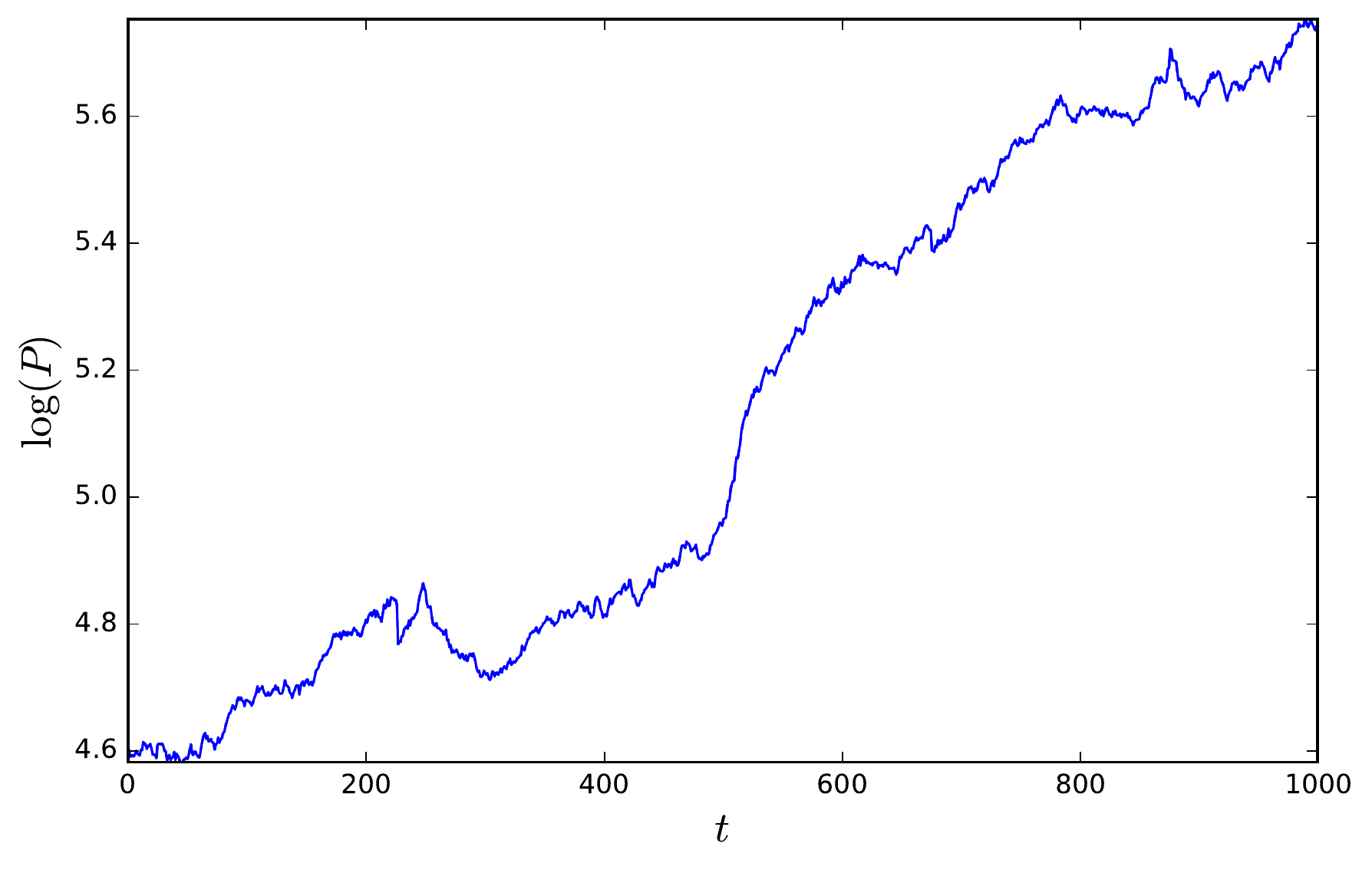}
  \includegraphics[width=8cm, height=5cm]{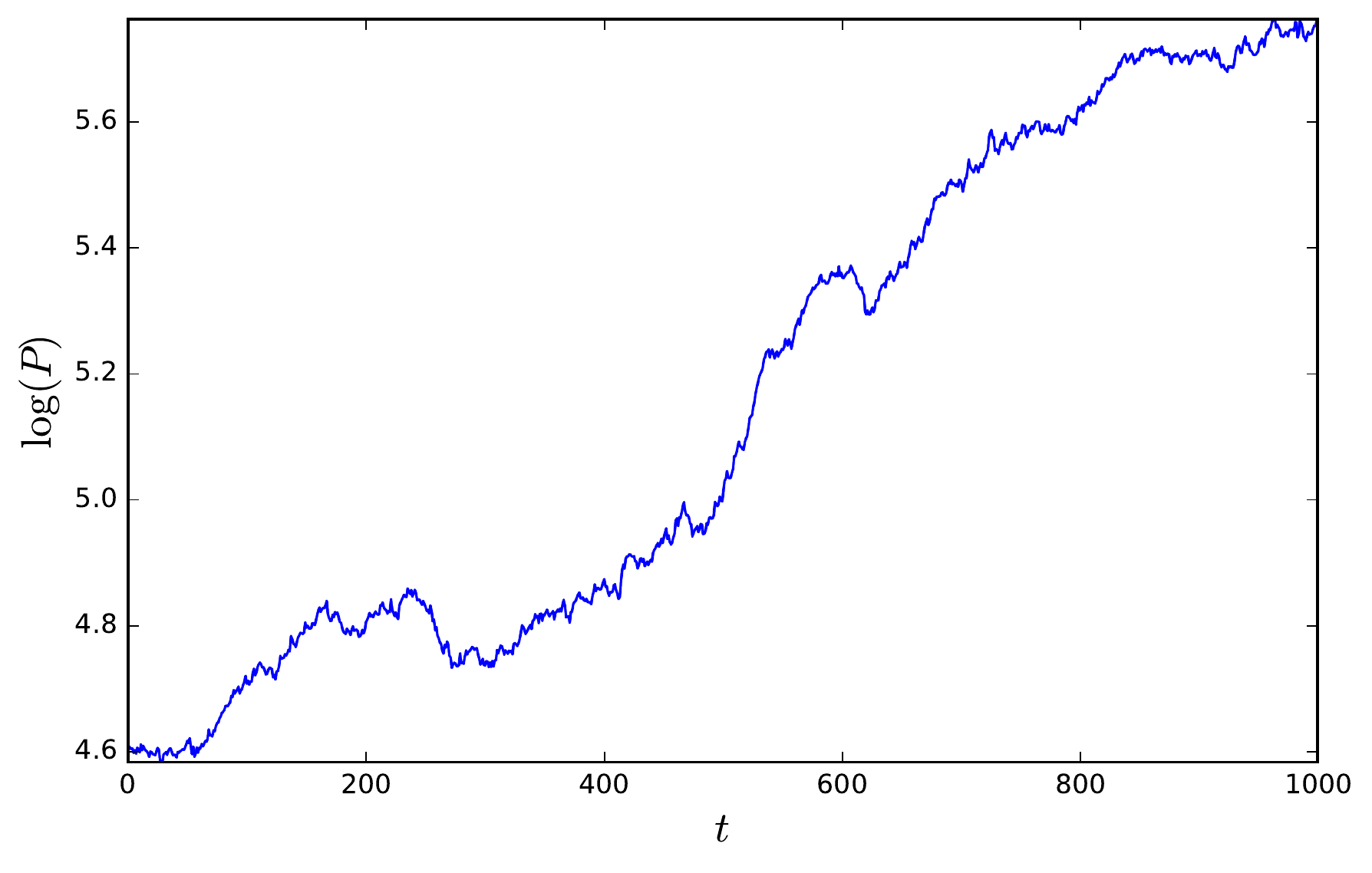}
   \includegraphics[width=8cm, height=5cm]{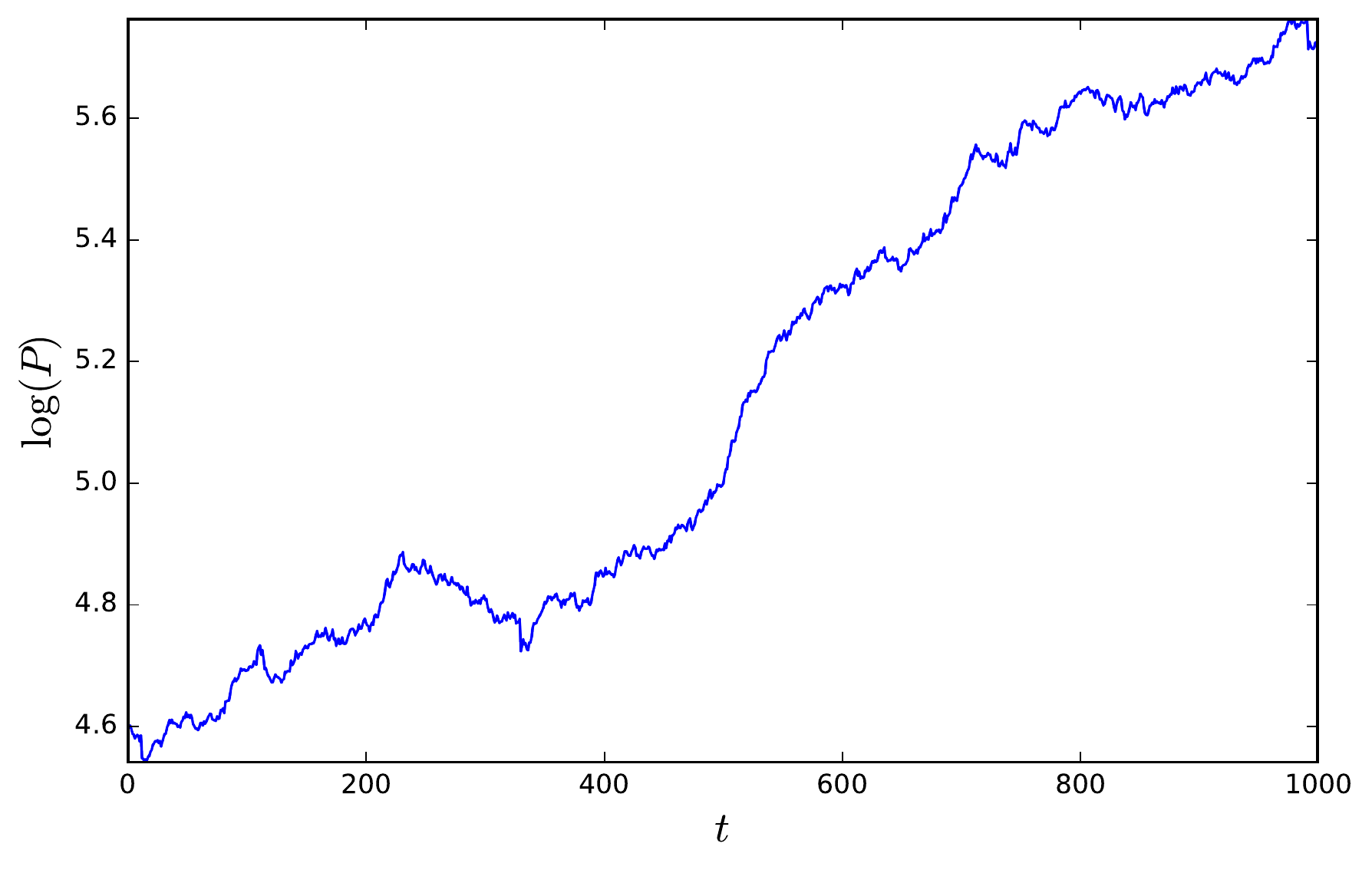}
    \includegraphics[width=8cm, height=5cm]{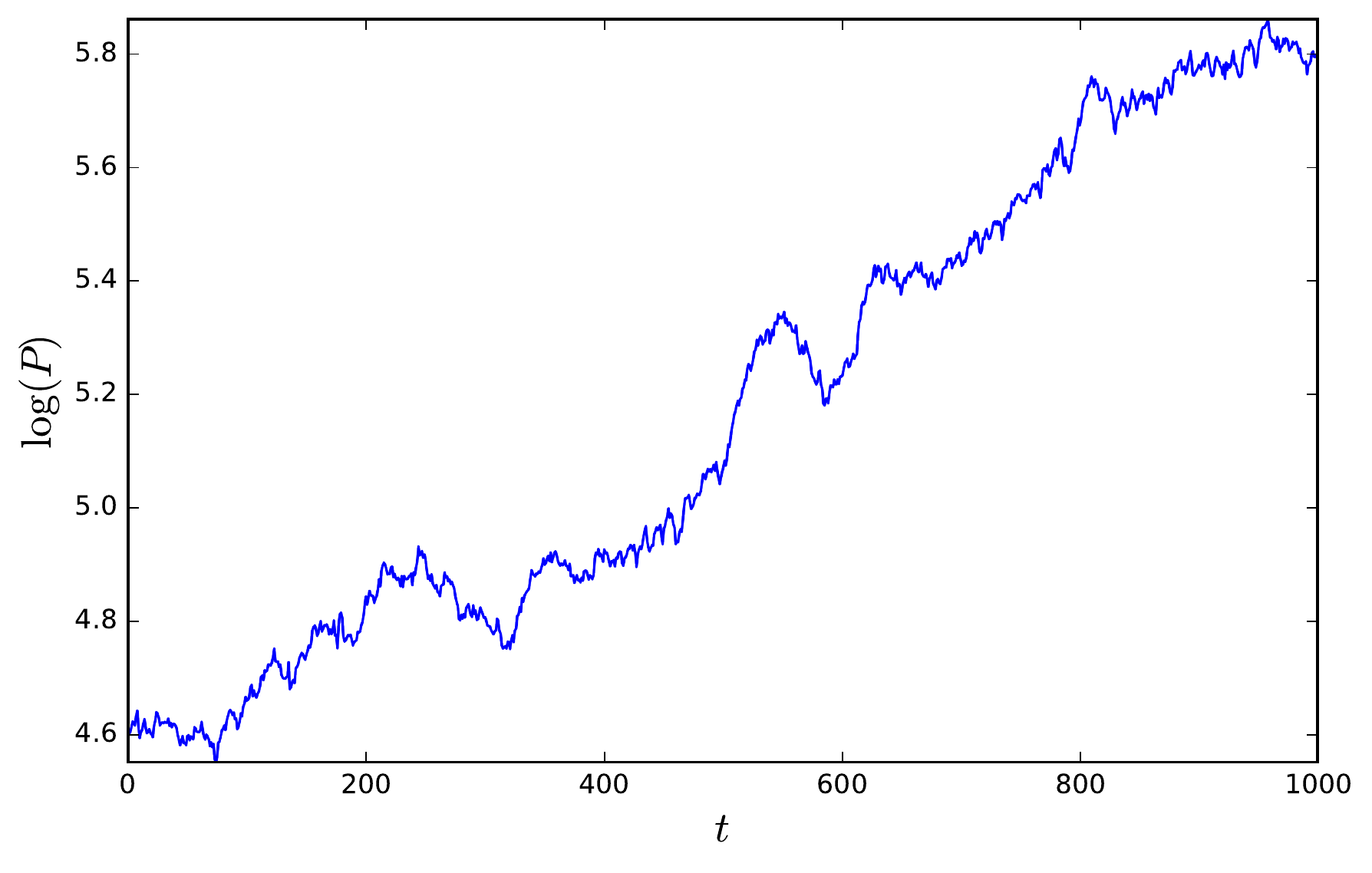}
 \caption{\label{stocks_section_6}  Time series of the logarithm of stock price for the initial
 (top to bottom, left to right) identical, uniform, normal, and Pareto wealth allocations
 in section \ref{section_6}.
}
\end{figure}

\begin{figure}
 \includegraphics[width=8cm, height=5cm]{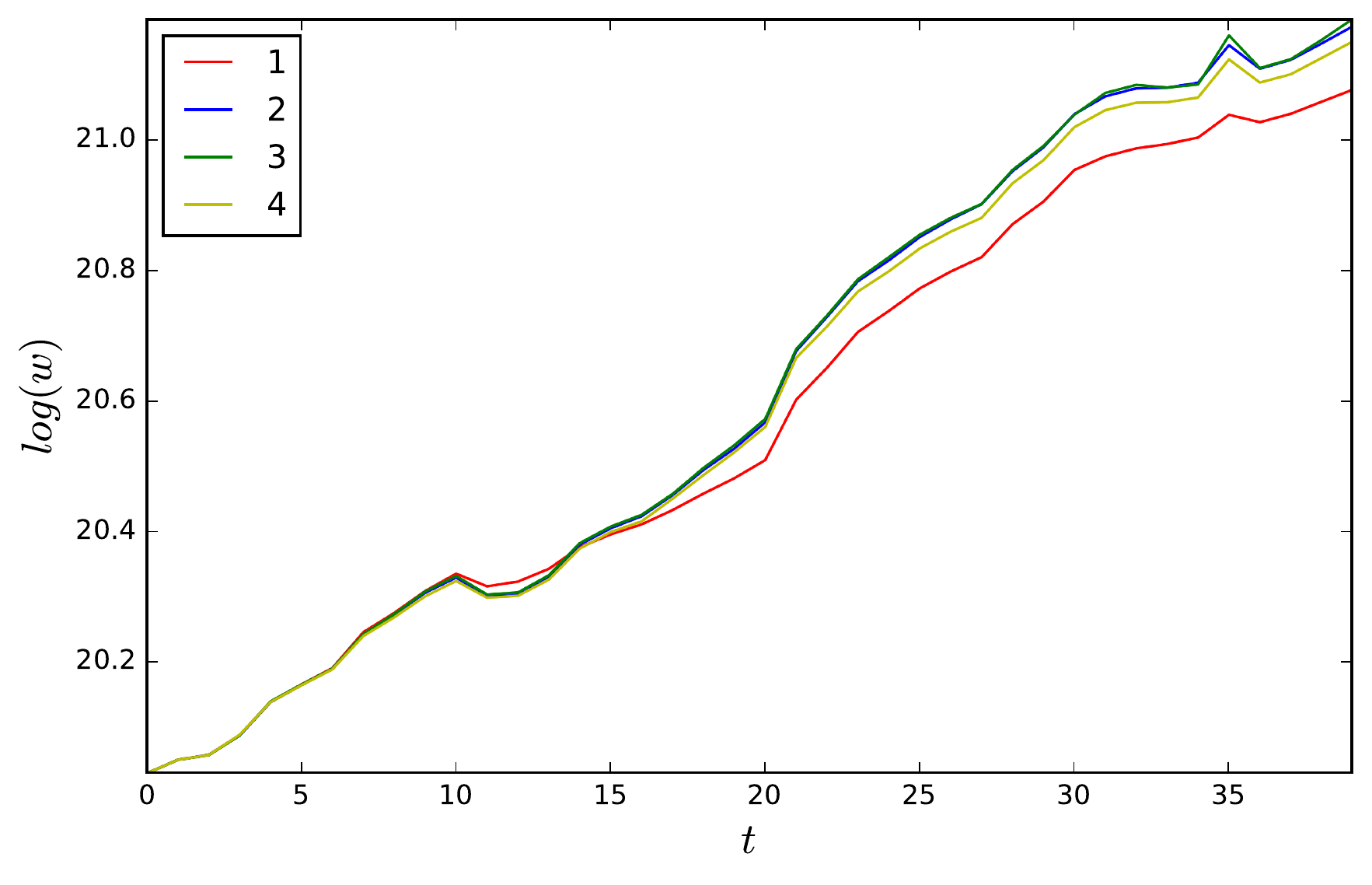}
  \includegraphics[width=8cm, height=5cm]{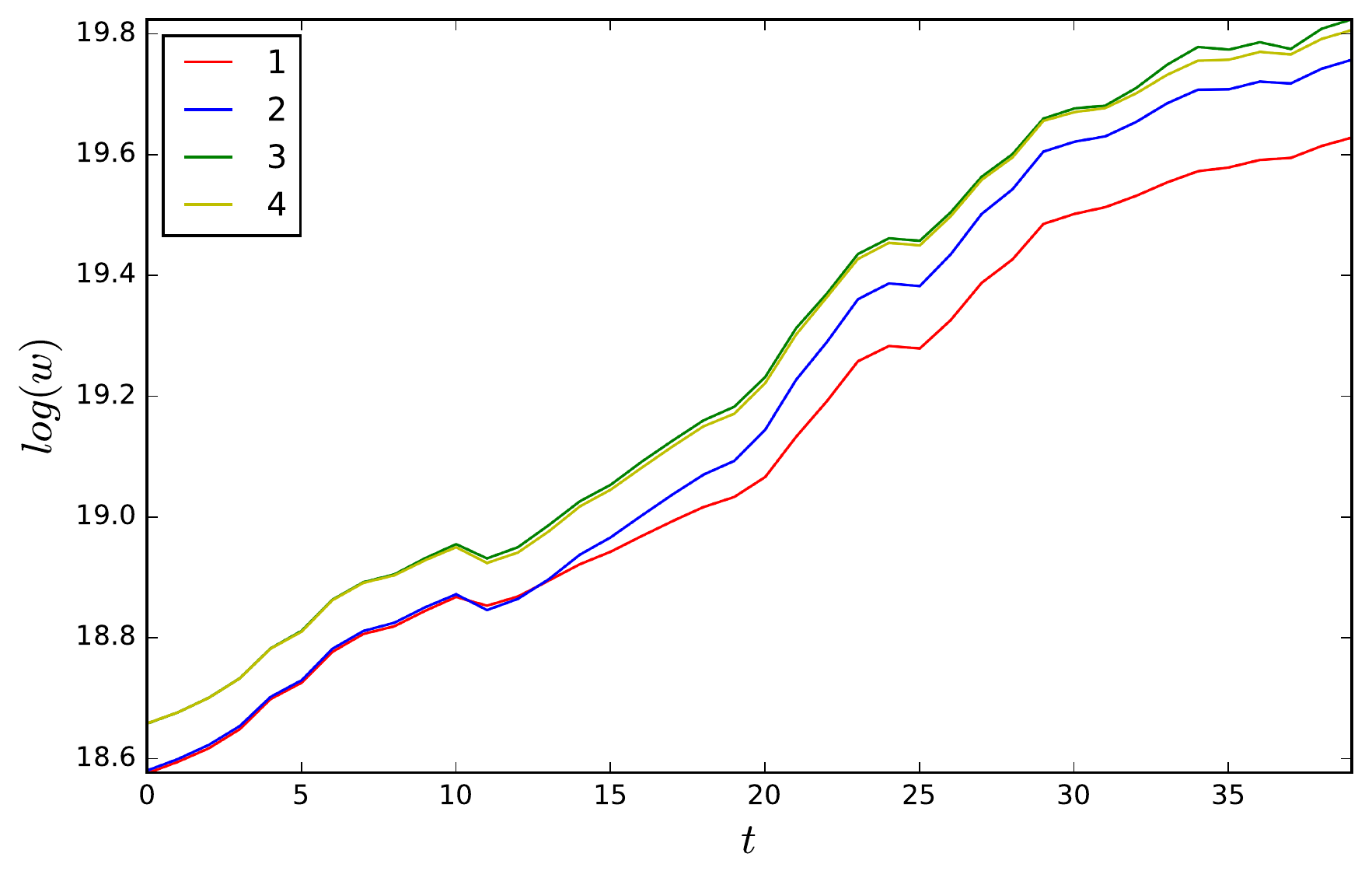}
   \includegraphics[width=8cm, height=5cm]{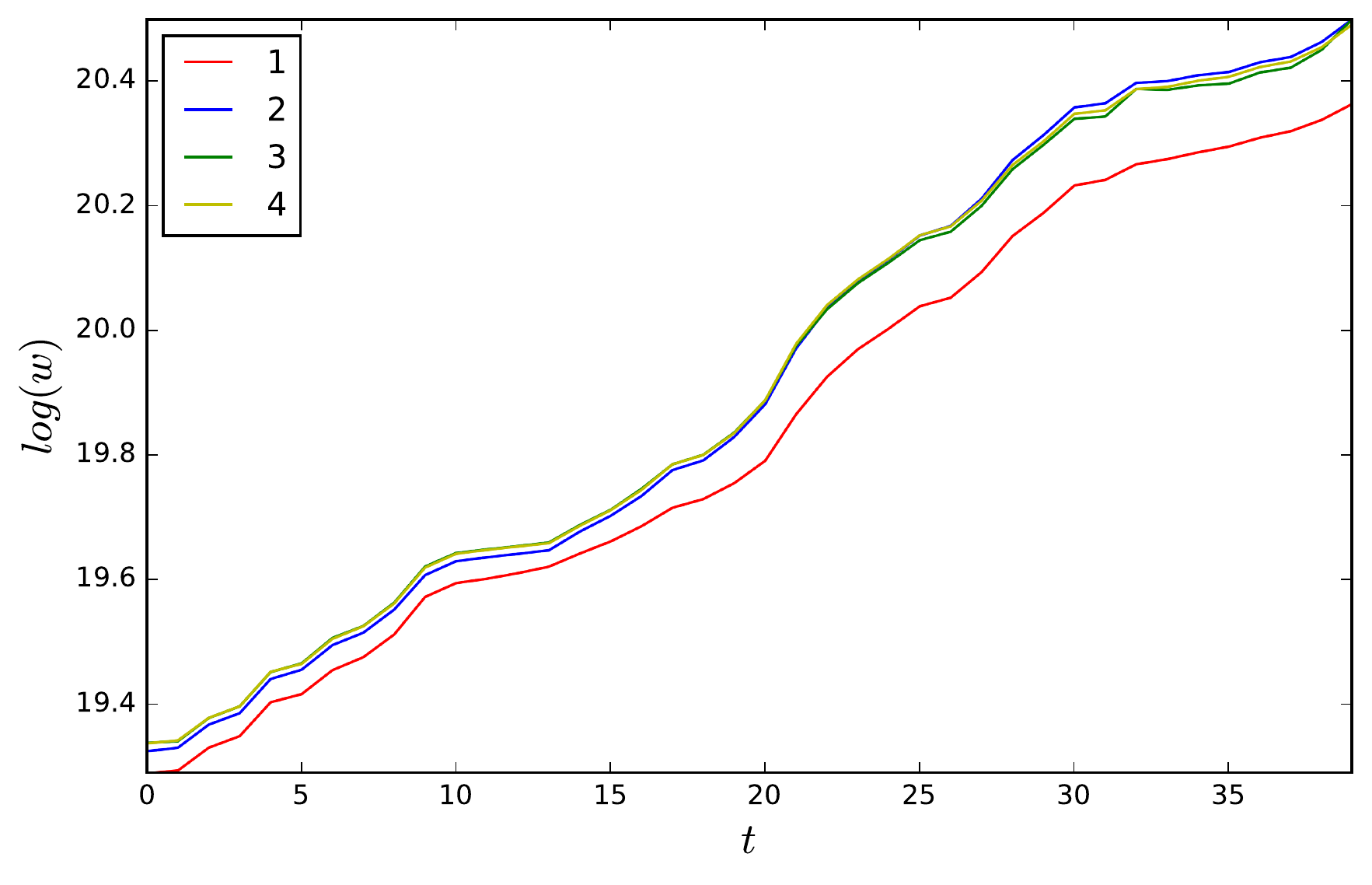}
    \includegraphics[width=8cm, height=5cm]{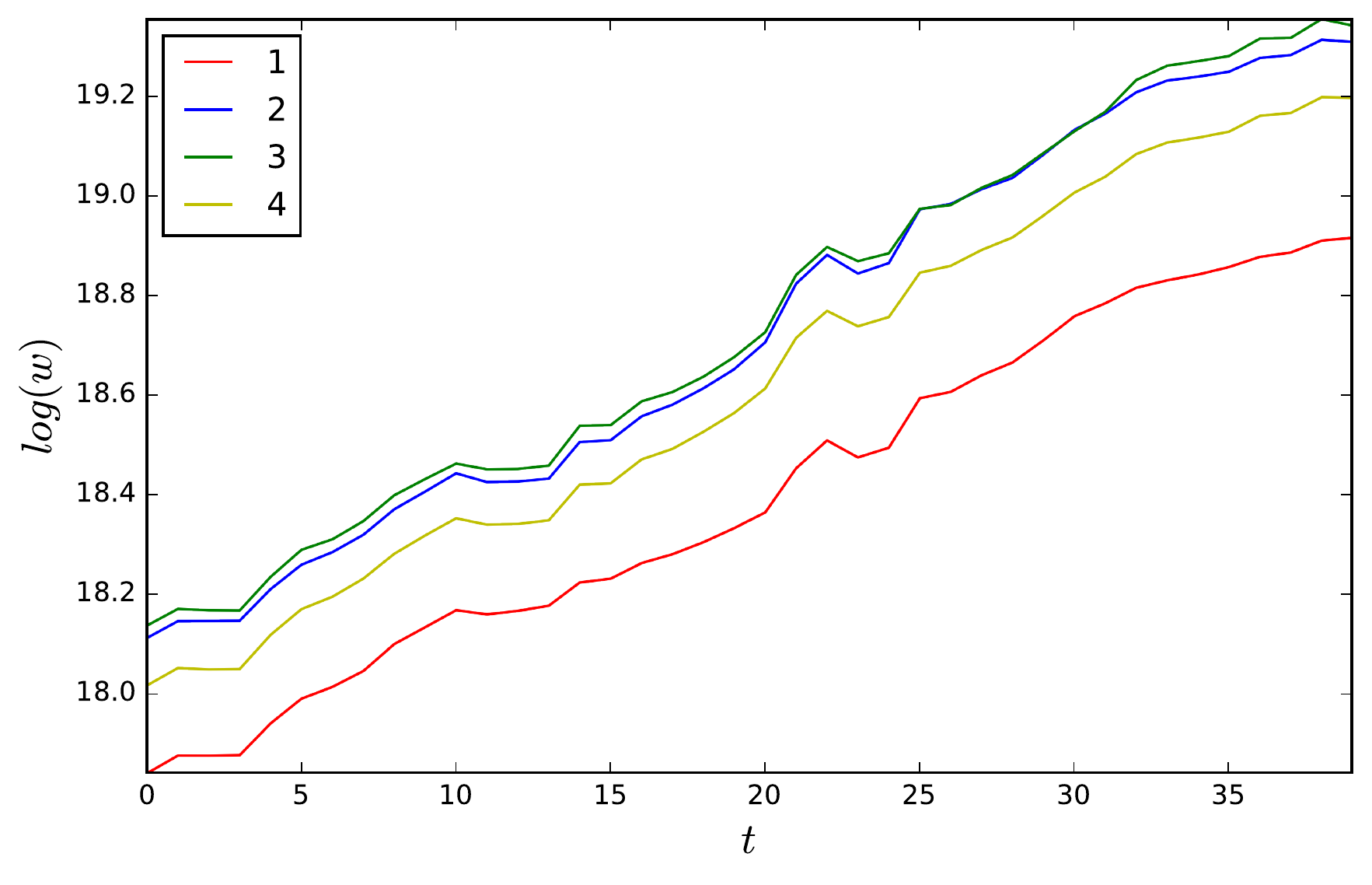}
 \caption{\label{group_wealth_6} Logarithms of total wealth of each of the four groups of agents for the initial
 (top to bottom, left to right) identical, uniform, normal, and Pareto wealth allocations
 in section \ref{section_6}. The measurements are taken every 25 time steps.
 The groups {\bf 1}, {\bf 2}, {\bf 3}, {\bf 4}, defined in section  \ref{section_6}
 by their different buy/sell sentiments $\psi$, are color-labeled.
}
\end{figure}

In section \ref{section_5} we saw how switching on a sell sentiment
experienced by all of the market participants results in a steep
decline of the stock price, readjusting it according to eq. (\ref{equilibrium_psi_price}).
It is natural to expect that if we activate the sell sentiment
among only a subgroup of the market participants the stock
decline will be more gradual.

In this section we consider the system of $N=1000$ agents trading over
$T=1000$ steps. We will fix the interest rate to be $r=5\times 10^{-4}$.
The dividend yield will be a non-trivial process
\begin{equation}
d = \left \{
  \begin{aligned}
    &{\cal N}(10^{-3},2\times 10^{-4}), && \text{if}\ t\in [0,3T/4] \\
    &0, && \text{if}\ t \in [3T/4,T]
  \end{aligned} \right.
\end{equation} 
We will assume that the calm and breaking news volatilities
are $\sigma_c={\cal N}(0.05,0.001)$, $\sigma_b={\cal N}(0.2,0.001)$,
where the breaking news is the Poisson process arriving
with the mean inverse time $\lambda^{-1}={\cal N}(0.005,0.0001)$.
The agents will participate according to the Poisson
process with the mean inverse arrival time $\rho^{-1}={\cal N}(0.6,0.01)$.

We will divide the agents into four groups with $250$ agents in each group.
The groups {\bf 1}, {\bf 2}, {\bf 3}, {\bf 4} will be fed the buy/sell sentiment process
\begin{align}
  \psi_1 &= \left \{
  \begin{aligned}
    &{\cal N}(0,0.1), && \text{if}\ t\in [0,T/4] \\
    &{\cal N}(-1,0.1), && \text{if}\ t \in [T/4,T/2]\\
    &{\cal N}(0,0.1), && \text{if}\ t \in [T/2,3T/4]\\
    &{\cal N}(-1,0.1), && \text{if}\ t \in [3T/4,T]
  \end{aligned} \right.\\
  \psi_2&= \left \{
  \begin{aligned}
    &{\cal N}(0,0.1), && \text{if}\ t\in [0,T/2] \\
    &{\cal N}(0.1,0.1), && \text{if}\ t \in [T/2,3T/4]\\
    &{\cal N}(0,0.1), && \text{if}\ t \in [3T/4,T]
  \end{aligned} \right.\\
  \psi_3 &= \left \{
  \begin{aligned}
    &{\cal N}(0,0.1), && \text{if}\ t\in [0,3T/4] \\
    &{\cal N}(1,0.1), && \text{if}\ t \in [3T/4,T]\\
  \end{aligned} \right.\\
  \psi_4 &= 
    {\cal N}(0,0.1), \quad \text{if}\ t\in [0,T] \,.
\end{align} 

The resulting stock evolution is plotted in fig. \ref{stocks_section_6}.
\footnote{Again notice that Pareto society is the most volatile \cite{RabertoCF2003},
in agreement with the results in section \ref{section_4} and section \ref{section_5}.}
We chose to plot the logarithm of stock price,
to visualize better the price change on top of the exponential inflation.
We can see a mild short-term decline starting at $t=T/4$ brought by the strong selling
preference of the group {\bf 1} during the period $[T/4,T/2]$. The stock
price also has a positive cusp at $t=T/2$, when the agents
of group {\bf 1} become buy/sell neutral again, and the agents
of group {\bf 2} develop a slight preference for buy over sell.
At $t=3T/4$ group {\bf 1} acquires a strong sell preference, and group {\bf 3}
acquires a strong buy preference, offsetting each other.
The stock price still has a slight negative cusp at $t=3T/4$,
due to vanishing dividend during the period $[3T/4,T]$,
thereby reducing the cash inflow into the system, and consequently
resulting in a lower slope of the stock inflation uptrend.

In fig. \ref{group_wealth_6} we plot time dependence of the logarithm of the collective wealth of
each of the four groups
of agents, recording the wealth every 25 steps. 
\footnote{Notice that the Pareto distribution, having for $a=1.5$ an infinite
variance, results in a noticeable spread in the initial allocation of wealth to 
the groups. This might position the group {\bf 1} off to a better start, somewhat offsetting
its subsequent poor trading behavior.
The group {\bf 3}, which has the best trading behavior in the considered market environment,
might end up worse than expected, if it is off at a disadvantaged Pareto start.
These phenomena can be observed in simulations.}
The base reference is provided by the group {\bf 4} (yellow line) whose agents
remained buy/sell neutral throughout the whole simulation. 
Before $t=T/4$ all the groups behave in a similar way:
everyone is following a neutral buy/sell sentiment during that period.

At  $t=T/4$ the group {\bf 1} (red line) starts preferring selling
over buying, therefore driving the stock price down. The agents of group {\bf 1}
are the least to suffer from the immediate stock decline, due to their relative pulling out of the stock market.
The groups {\bf 2}, {\bf 3}, {\bf 4} are buy/sell neutral during that
period, and their wealth evolves in a synchronous manner (as is best seen
on the plot corresponding to the identical initial wealth allocation).
However even before $t=T/2$ the negative gap develops between the wealth of the group {\bf 1}
and the wealth of the groups {\bf 2}, {\bf 3}, {\bf 4}. This is due to the fact that the
 groups {\bf 2}, {\bf 3}, {\bf 4}, possessing
more shares of stock than the group {\bf 1}, have accumulated more of the dividend returns on the stocks.

At $t=T/2$ the group {\bf 1} becomes buy/sell neutral again, and starts acquiring more
shares of stock. This, and the slight buy preference of the group {\bf 2} (blue line)
during the time $[T/2,3T/4]$, results in a positive wealth cusp at $t=T/2$.
The group {\bf 2}
being the most invested, is expected to benefit the most from the stock price uptrend.

At $t=3T/4$ the group {\bf 1} starts actively pulling out of the stock market again
and the group {\bf 3} (green line) starts actively buying into stocks.
The stock does not issue dividends during the time interval $[3T/4,T]$.
Notice that the activity of the group {\bf 3} offsets the activity of group {\bf 1}.
(We have
confirmed this by running a similar simulation with identically vanishing $r$ and $d$,
where it is clear to see if price trend develops, transitioning to a new equilibrium value.)
Notice that since the stock does not pay dividend during the period $[3T/4,T]$,
the money inflow into the system decreases, and as a result the stock price
uptrend is only fueled by the cash interest payments received by the agents.
Therefore the stock price, as well as the agents's portfolios, have a smaller
slope during that period, and a negative cusp at $t=3T/4$.

\section{Discussion}\label{discussion}

In this paper we proposed and studied a simulated stock market
environment with the sentiment drivers determining trading
actions of participating agents. We have considered
sentiments affecting the trading intensity, buy/sell preference,
and stock price volatility. We have omitted entirely
any strategy assignment to the agents. In our models
the stock price behavior, observed during the simulation, does not influence the trading
decisions made by the agents. This is in contrast with most
of the literature dedicated to the artificial stock markets.

Our paper is to be applied to the study of stock market
dynamics after the question of the market participants's trading
behavior has already been answered. We assume that the answer
to that question boils down to specifying the sentiment processes followed by a large groups of
the market participants.
That is, our study of the stock dynamics is done in the framework
where the market state is in one to one correspondence with the
set of sentiment driving processes.
Once existence of those sentiment
processes is confirmed and their parameters are established,
our paper can be used to determine the resulting stock
price dynamics. 
The assumption that the agents 
will follow the sentiment after observing the stock price
behavior emergent in the course of simulation
is rather strong, and is used as a postulate for our sentiment-driven market framework.

We have studied several simulations in our proposed sentiment-driven
market environment. We have noticed that, as expected, incorporation
of a non-trivial volatility sentiment process results
in a deviation of a large stock returns
from the log-normal distribution.
This is consistent with the actual observations of the statistical
properties of stock price returns, and is typically
sought to be reproduced in the stock market simulations.
We have also demonstrated how a non-trivial buy/sell sentiment
process creates a predictable price trends, and allows to manipulate
the price dynamics away from a simple mean-reverting 
behavior. In particular we have studied the situation in which different
subgroups of agents are influenced by different sentiments while trading
with each other freely.

We have considered four possible kinds of initial wealth
allocation: identical, uniform, normal, and Pareto,
and studied the subsequent wealth dynamics of all and subgroups
of the agents over the course of simulation. It is interesting that regardless
of the considered initial wealth allocation, and consistent with the literature
on this subject, the wealth distribution of the agents quickly acquires
a power-law Pareto tail. We have studied how the Pareto
exponent evolves in time.

A possible application of our model
would be to use it as an ingredient of a trading strategy,
built on the assumption that one
can derive the properties of the stock market behavior,
such as price trends/volatility and trade volume, by knowing
what sentiment processes that behavior can be reduced to.
Then observing the stock price time dependence we can
use Bayesian inference to obtain the posterior likelihoods
of the sentiment processes. This will allows us to have some
idea of where so derived sentiments will lead the stock price.

\section*{Acknowledgements} This work was supported by the Oehme Fellowship.
I would like to thank I.~Teimouri for useful comments on the draft of this paper.

\section{Appendix: The stock exchange structure}\label{The_matching_engine}

In this appendix we describe design of
our stock exchange.
We have implemented all of our calculations in Python.
The stock exchange is an interaction mediator for the agents.
Each agent is characterized by a portfolio: the number of shares
of stock, and the quantity of cash.
Information about portfolios of agents is kept in the client book,
which we describe in subsection \ref{The_client_book_specifics}.
All of the market activity of agents is done through the intermediary
of the stock exchange. At each time step the stock exchange begins
by accepting orders from the agents.
It records orders into the sorted tables of the order book,
which we describe in subsection \ref{The_order_book_specifics}.
Then it switches on the
matching engine which tries to find a new equilibrium price
by balancing the supply and demand for the shares of stock.
If the equilibrium is reached, the matching engine fills all of the possible
orders, as described in subsection \ref{The_matching_engine_specifics}.
It then interfaces with the client book, and updates portfolios of the agents
accordingly to the filled orders.

\subsection{The client book}\label{The_client_book_specifics}

The purpose of the client book is to maintain the portfolio records of all of the agents,
with each agent being assigned a unique client ID number.
We implement the client book as a class which has an easy interface, allowing to add new clients,
endow clients with the shares of stock and units of cash, modify client
records, and delete clients. In our system each client will be allowed to maintain
only one open order at a time. As described in the next subsection each order is identified
by a unique order ID number. The client book maintains a table matching the order ID
to the ID of the client who issued that order.

\subsection{The order book}\label{The_order_book_specifics}

All of the orders from the agents go directly to the order book. The orders are characterized
by the order ID number, the side (buy or sell), the limit price, and the size.
We implement the order book as a class, which maintains separate dictionaries of buy and sell
orders, matching the sorted prices to the sizes and identification numbers of the orders.
The order book provides an easy interface allowing to add, modify,
or cancel an order. If the stock exchange fills only a part of the order (as it frequently
occurs for the orders at the price level equal to the exact intersection price
of the supply and demand curves, see the next subsection) it will modify
the corresponding order to the smaller size accordingly.

\subsection{The matching engine}\label{The_matching_engine_specifics}

The center of the stock exchange is the matching engine. In our implementation the matching engine
is a class which inherits from the client book and the order book classes,
allowing it to interface directly with the client records and the order records.
The matching engine has two main functionalities. The first is to match the orders of the order book,
and determine the equilibrium price for the supply and demand. The second is to
fill the orders, in accord with the matched price, and update the client book correspondingly. 

The matching of orders in the order book is done by 
constructing the supply and demand curves on the price and volume plane and looking
for intersection. The supply at price $P$ is given
by all the sell orders at price $P$ and lower,
and is therefore a non-decreasing function of price.
The demand at price $P$ is given by all the buy
orders at price $P$ and higher, and is therefore a non-increasing
function of price. Both curves are discrete, and therefore resovling
their intersection requires a careful consideration.

First of all, it might be that the intersection is never achieved.
If the highest buy order price $P_1$ is strictly lower than the
lowest sell order price $P_2$, then no orders can be filled. This is illustrated
by the following figure.

\setlength{\unitlength}{1.5cm}
\begin{picture}(5,5)
\multiput(0,0.02)(0.4,0){13}{\line(1,0){0.3}}
\multiput(5.2,0)(0,0.4){3}{\line(0,1){0.3}}
\multiput(5.2,1.2)(0.4,0){3}{\line(1,0){0.3}}
\multiput(6.4,1.2)(0,0.4){5}{\line(0,1){0.3}}
\multiput(6.4,3.1)(0.4,0){2}{\line(1,0){0.3}}
\multiput(7.2,3.1)(0,0.4){3}{\line(0,1){0.3}}
\multiput(7.2,4.3)(0.4,0){2}{\line(1,0){0.3}}
\put(8,0){\line(-1,0){6}}
\put(2,0){\line(0,1){2}}
\put(2,2){\line(-1,0){1.5}}
\put(0.5,2){\line(0,1){2}}
\put(0.5,4){\line(-1,0){0.5}}
\put(1.9,-0.35){$P_1$}
\put(5.1,-0.35){$P_2$}
\put(0,0.1){\vector(1,0){0.5}}
\put(0.5,0.2){$P$}
\put(0,0.1){\vector(0,1){0.5}}
\put(0.1,0.5){$V$}
\end{picture}
\vspace{1cm}

We are plotting the supply curve (made up by the sell orders)
using the punctured line, and the demand curve (made up
by the buy orders) using the solid line. The horizontal axis
is the price, the vertical axis is the volume of the orders, the base line
represents zero volume.
The supply curve (punctured line) starts at zero volume
at zero price (no one wants to sell at zero price)
and jumps up by $V_s$ each time we cross a price point $P_s$ of some sell order
of the size $V_s$.
The supply order saturates to the total volume of all the sell
orders at the highest submitted sell order price, where everyone would
be willing to sell.
The demand curve (solid line) starts at the total demand volume
(of all the buy orders combined) at zero price (everyone wants to buy at zero
price). As the price goes up, each time we cross a price point $P_b$
of some buy order of the size $V_b$, the demand curve drops by $V_b$.
After the highest submitted buy price price the demand is zero: no one wants to 
buy at the higher price.

If the intersection occurs, it can be of three possible types.
The type of intersection determines the intersection price and
volume, and resolves how the orders at the intersection price are filled.
We categorize the intersection types according to what curve (supply or/and demand)
is vertical at the intersection point.

\begin{itemize}

\item Buy cross. The demand curve is vertical at the intersection point.

\setlength{\unitlength}{1.5cm}
\begin{picture}(5,5)
\multiput(0,0)(0.4,0){3}{\line(1,0){0.3}}
\multiput(1.2,0)(0,0.4){3}{\line(0,1){0.3}}
\multiput(1.2,1.2)(0.4,0){3}{\line(1,0){0.3}}
\multiput(2.4,1.2)(0,0.4){3}{\line(0,1){0.3}}
\multiput(2.4,2.4)(0.4,0){7}{\line(1,0){0.3}}
\multiput(5.2,2.4)(0,0.4){5}{\line(0,1){0.3}}
\multiput(5.2,4.3)(0.4,0){4}{\line(1,0){0.3}}
\put(7,0){\line(-1,0){1}}
\put(6,0){\line(0,1){1.75}}
\put(6,1.75){\line(-1,0){2}}
\put(4,1.75){\line(0,1){1.5}}
\put(4,3.25){\line(-1,0){3.5}}
\put(0.5,3.25){\line(0,1){1}}
\put(0.5,4.25){\line(-1,0){0.5}}
\put(4.2,1.9){\vector(0,1){0.35}}
\put(4.25,2){$\Delta$}
\put(4,1.3){\vector(0,1){0.35}}
\put(3.9,1){$P_*$}
\put(1.95,2.4){\vector(1,0){0.35}}
\put(1.56,2.35){$V_*$}
\put(0,0.1){\vector(1,0){0.5}}
\put(0.5,0.2){$P$}
\put(0,0.1){\vector(0,1){0.5}}
\put(0.1,0.5){$V$}
\end{picture}

In this case all the sell orders at prices lower than $P_*$ can be filled,
and the resulting volume $V_*$ is defined by the cumulative sell
volume at $P_*$. We start filling the buy
orders at the highest submitted buy price. We can first fill all those
orders for the prices strictly higher than $P_*$. After this is achieved
we will have $\Delta$ less buy orders filled than the sell orders.
The $\Delta$ is in general a part of all the buy orders
submitted at the price $P_*$. We fill them in the order of arrival (as determined by the
order ID's),
and keep the rest of the orders in the order book.

\item Sell cross. The supply curve is vertical at the intersection point.
Symmetric to the buy cross w.r.t. the buy/sell exchange.

\setlength{\unitlength}{1.5cm}
\begin{picture}(5,5)
\multiput(0,0)(0.4,0){3}{\line(1,0){0.3}}
\multiput(1.2,0)(0,0.4){3}{\line(0,1){0.3}}
\multiput(1.2,1.2)(0.4,0){3}{\line(1,0){0.3}}
\multiput(2.4,1.2)(0,0.4){5}{\line(0,1){0.3}}
\multiput(2.4,3.1)(0.4,0){7}{\line(1,0){0.3}}
\multiput(5.2,3.1)(0,0.4){3}{\line(0,1){0.3}}
\multiput(5.2,4.3)(0.4,0){4}{\line(1,0){0.3}}
\put(7,0){\line(-1,0){1}}
\put(6,0){\line(0,1){2}}
\put(6,2){\line(-1,0){5.5}}
\put(0.5,2){\line(0,1){2}}
\put(0.5,4){\line(-1,0){0.5}}
\put(2.2,1.5){\vector(0,1){0.35}}
\put(1.9,1.6){$\Delta$}
\put(2.4,0.7){\vector(0,1){0.35}}
\put(2.3,0.35){$P_*$}
\put(0,2){\vector(1,0){0.35}}
\put(-0.3,2){$V_*$}
\put(0,0.1){\vector(1,0){0.5}}
\put(0.5,0.2){$P$}
\put(0,0.1){\vector(0,1){0.5}}
\put(0.1,0.5){$V$}
\end{picture}

In this case all the buy orders at prices higher than $P_*$ can be filled,
and the resulting volume $V_*$ is defined by the cumulative buy
volume at $P_*$. We start filling the sell
orders at the lowest submitted sell price. We can first fill all those
orders for the prices strictly lower than $P_*$. After this is achieved
we will have $\Delta$ less sell orders filled than the buy orders.
The $\Delta$ is in general a part of all the sell orders
submitted at the price $P_*$. We fill them in the order of arrival,
and keep the rest of the orders in the order book.

\item Mixed cross. Both the supply and the demand curves are vertical at the intersection point.
That is, there are buy and sell orders submitted at the price $P_*$. 
Depending on the buy and sell volumes at price $P_*$
there are two cases.
In the first case we can fill
all the sell orders and the fraction $\Delta$ of the buy orders.

\setlength{\unitlength}{1.5cm}
\begin{picture}(5,5)
\multiput(0,0)(0.4,0){3}{\line(1,0){0.3}}
\multiput(1.2,0)(0,0.4){3}{\line(0,1){0.3}}
\multiput(1.2,1.2)(0.4,0){7}{\line(1,0){0.3}}
\multiput(3.95,1.2)(0,0.4){3}{\line(0,1){0.3}}
\multiput(4,2.3)(0.4,0){3}{\line(1,0){0.3}}
\multiput(5.2,2.4)(0,0.4){5}{\line(0,1){0.3}}
\multiput(5.2,4.3)(0.4,0){4}{\line(1,0){0.3}}
\put(7,0){\line(-1,0){1}}
\put(6,0){\line(0,1){1.75}}
\put(6,1.75){\line(-1,0){2}}
\put(4,1.75){\line(0,1){1.5}}
\put(4,3.25){\line(-1,0){3.5}}
\put(0.5,3.25){\line(0,1){1}}
\put(0.5,4.25){\line(-1,0){0.5}}
\put(4.2,1.9){\vector(0,1){0.35}}
\put(4.25,2){$\Delta$}
\put(4,0.8){\vector(0,1){0.35}}
\put(3.9,0.5){$P_*$}
\put(3.5,2.3){\vector(1,0){0.35}}
\put(3.1,2.2){$V_*$}
\put(0,0.1){\vector(1,0){0.5}}
\put(0.5,0.2){$P$}
\put(0,0.1){\vector(0,1){0.5}}
\put(0.1,0.5){$V$}
\end{picture}

In the second case we can fill all the buy
orders and the fraction $\Delta$ of the sell orders.

\setlength{\unitlength}{1.5cm}
\begin{picture}(5,5)
\multiput(0,0)(0.4,0){3}{\line(1,0){0.3}}
\multiput(1.2,0)(0,0.4){3}{\line(0,1){0.3}}
\multiput(1.2,1.2)(0.4,0){7}{\line(1,0){0.3}}
\multiput(3.95,1.2)(0,0.4){7}{\line(0,1){0.3}}
\multiput(4,3.9)(0.4,0){3}{\line(1,0){0.3}}
\multiput(5.2,3.9)(0,0.4){1}{\line(0,1){0.3}}
\multiput(5.2,4.3)(0.4,0){4}{\line(1,0){0.3}}
\put(7,0){\line(-1,0){1}}
\put(6,0){\line(0,1){1.75}}
\put(6,1.75){\line(-1,0){2}}
\put(4,1.75){\line(0,1){1.5}}
\put(4,3.25){\line(-1,0){3.5}}
\put(0.5,3.25){\line(0,1){1}}
\put(0.5,4.25){\line(-1,0){0.5}}
\put(3.8,1.9){\vector(0,1){1.2}}
\put(3.45,2.4){$\Delta$}
\put(4,0.8){\vector(0,1){0.35}}
\put(3.9,0.5){$P_*$}
\put(4.5,3.25){\vector(-1,0){0.35}}
\put(4.7,3.2){$V_*$}
\put(0,0.1){\vector(1,0){0.5}}
\put(0.5,0.2){$P$}
\put(0,0.1){\vector(0,1){0.5}}
\put(0.1,0.5){$V$}
\end{picture}

The exact match is a particular case of the mixed cross with $\Delta=0$.

\end{itemize}

After the possible orders are filled, we can either keep
the remaining orders in the order book pending, or clear up
the order book.

\end{document}